\documentclass[fleqn,usenatbib]{mnras}

\usepackage{newtxtext,newtxmath}

\usepackage[T1]{fontenc}

\DeclareRobustCommand{\VAN}[3]{#2}
\let\VANthebibliography\thebibliography
\def\thebibliography{\DeclareRobustCommand{\VAN}[3]{##3}\VANthebibliography}

\usepackage{graphicx}	
\usepackage{amsmath}	
\usepackage{pdflscape} 
\usepackage{booktabs} 
\usepackage{lipsum}  
\usepackage{layout}
\usepackage{xprintlen}
\usepackage{tabularx} 
\usepackage{hyperref} 
\usepackage{enumerate} 

\usepackage{xcolor} 

\providecommand{\e}[1]{\ensuremath{\times 10^{#1}}}
\usepackage{siunitx}
\usepackage{fancyvrb}
\usepackage{xspace}
\usepackage[normalem]{ulem}

\usepackage{newtxtext,newtxmath}

\defcitealias{mahlkeAsteroidPhaseCurves2021}{M21}
\defcitealias{schemelZwickyTransientFacility2021}{SB21}
\defcitealias{mcneillComparisonPhysicalProperties2021}{McN21}
\defcitealias{bowellApplicationPhotometricModels1989}{B89}
\defcitealias{penttilaG1G2Photometric2016}{P16}

\title[ATLAS phase curves]{Main-belt and Trojan Asteroid Phase Curves from the ATLAS Survey}

\author[J. E. Robinson et al.]{
	James E. Robinson,$^{1}$\thanks{E-mail: james.robinson@ed.ac.uk} 
	 Alan Fitzsimmons,$^{2}$ 
	 David R. Young,$^{2}$ 
	 Michele Bannister,$^{3}$ 
     Larry Denneau,$^{4}$ 
     \newauthor
     Nicolas Erasmus, $^{5}$ 
  	 Amanda Lawrence,$^{4}$ 
	 Robert J. Siverd,$^{4}$ 
     John Tonry,$^{4}$ 
	\\
	$^{1}$Institute for Astronomy, The University of Edinburgh, Blackford Hill, Edinburgh EH9 3HJ, UK\\	
 $^{2}$Astrophysics Research Centre, Queen's University Belfast, University Road, BT7 1NN, Belfast, Northern Ireland\\
	 $^{3}$School of Physical and Chemical Sciences – Te Kura Mat\={u}, University of Canterbury, Christchurch 8041, New Zealand \\
	 	 $^{4}$ Institute for Astronomy, University of Hawaii, Honolulu, HI 96822, USA \\
	 $^{5}$ South African Astronomical Observatory, Cape Town 7925, South Africa
}

\date{Accepted 2024 April 4. Received 2024 March 22; in original form 2023 July 28}

\pubyear{2024}

\begin{document}
\label{firstpage}
\pagerange{\pageref{firstpage}--\pageref{lastpage}}
\maketitle

\newcommand{\ATLAS}{\mbox{ATLAS}\xspace}
\newcommand{\rockAtlas}{\textsc{rockAtlas}\xspace}
\newcommand{\pyephem}{\textsc{pyephem}\xspace}
\newcommand{\astorb}{\textsc{astorb}\xspace}
\newcommand{\orbfit}{\textsc{orbfit}\xspace}
\newcommand{\MPChecker}{\textsc{MPChecker}\xspace}
\newcommand{\dophot}{\textsc{dophot}\xspace}
\newcommand{\astdys}{\textsc{astdys}\xspace}
\newcommand{\healpix}{\textsc{HEALpix}\xspace}
\newcommand{\astropy}{\textsc{astropy}\xspace}
\newcommand{\sbpy}{\textsc{sbpy}\xspace}
\newcommand{\ssodnet}{\textsc{SsODNet}\xspace}

\begin{abstract}
Sparse and serendipitous asteroid photometry obtained by wide field surveys such as the Asteroid Terrestrial-impact Last Alert System (\ATLAS) is a valuable resource for studying the properties of large numbers of small Solar System bodies.
We have gathered a large database of \ATLAS photometry in wideband optical cyan and orange filters, consisting of 9.6\e{7} observations of 4.5\e{5} main belt asteroids and Jupiter Trojans.
We conduct a phase curve analysis of these asteroids considering each apparition separately, allowing us to accurately reject outlying observations and to remove apparitions and asteroids not suitable for phase curve determination.
We obtain a dataset of absolute magnitudes and phase parameters for over 100,000 selected asteroids observed by \ATLAS, $\sim66,000$ of which had sufficient measurements to derive colours in the \ATLAS filters.
To demonstrate the power of our dataset we consider the properties of the Nysa-Polana complex, for which the \ATLAS colours and phase parameters trace the S-like and C-like compositions amongst family members. 
We also compare the properties of the leading and trailing groups of Jupiter Trojans, finding no significant differences in their phase parameters or colours as measured by \ATLAS, supporting the consensus that these groups were captured from a common source population during planetary migration.
Furthermore, we identify $\sim9000$ asteroids that exhibit large shifts in derived absolute magnitude between apparitions, indicating that these objects have both elongated shapes and spin axes with obliquity $\sim 90\ \si{degrees}$.
\end{abstract}




\section{Introduction}\label{sec:intro}

%

There are currently over $10^6$ main belt asteroids (MBAs) located between the orbits of Mars and Jupiter that have been discovered and catalogued through astrometric observations. 
Once catalogued, their orbits are known and their location within the Solar System relative to the Sun and an Earth-based observer can be calculated for any given time. 
Measuring the apparent magnitude, $m_\lambda$, through photometric methods at a specific point in time then allows for an estimate of the absolute magnitude, $H_\lambda$, by fitting those observations to the formula
\begin{equation}
	H_\lambda=m_\lambda-5\log\left( R \Delta \right)-g_\lambda(\alpha)
	\label{eqn:absolute_magnitude}
\end{equation}
Here $R$ and $\Delta$ are the heliocentric and geocentric distances in au and $\alpha$ is the Sun-asteroid-Earth phase angle, all calculated at the time $m_\lambda$ was measured.
The phase function, $g_\lambda(\alpha)$, mathematically describes the macroscopic scattering law of sunlight from the asteroidal surface as a function of $\alpha$. For all models of asteroid light scattering, $g_\lambda(\alpha)$ behaves such that increasing phase angles leads to decreasing brightness in a monotonic fashion.
Furthermore, as phase angle approaches zero the  brightness of asteroids often increases rapidly in an opposition surge. 
The main drivers of this effect are believed to be the mechanisms of shadow hiding and coherent backscatter \citep[e.g.][]{muinonenAsteroidPhotometricPolarimetric2002,muinonenThreeparameterMagnitudePhase2010a,penttilaG1G2Photometric2016}.
Therefore by definition the absolute magnitude, $H_\lambda$, of a spherical asteroid is its magnitude when observed at $1\, \si{au}$ from both the Sun and the Earth, whilst at opposition ($\alpha = 0^\circ$).
In reality most asteroids are non-spherical, and so the measured $H_\lambda$ will also depend on its orientation as viewed by the observer.

Due to the large number of known asteroids and the commensurate difficulty of obtaining physical or spectral information for individual objects, the measurement of $H_\lambda$ and $g_\lambda(\alpha)$ often provides the first estimate of asteroid properties for a large fraction of the population. 
Even then, these parameters may have significant uncertainties; for instance if the photometric measurements that are used to derive $g_\lambda(\alpha)$ only cover a small range of phase angles the phase function may not be well constrained.
Most photometric measurements are currently reported by large wide-field sky surveys such as Pan-STARRS \citep[Panoramic Survey Telescope and Rapid Response System;][]{kaiserPanSTARRSWidefieldOptical2010}, the Catalina and Mount Lemmon Sky Survey \citep{christensenCatalinaSkySurvey2012}, and \ATLAS \citep[Asteroid Terrestrial-impact Last Alert System;][]{tonryATLASHighcadenceAllsky2018}. 
All of these surveys use different filters, which means the reported $m_\lambda$ requires transformation to a common standard filter, normally Johnson $V$-band for historical reasons. Additionally, the wide range of spectral types of asteroids \citep{demeoExtensionBusAsteroid2009} means the correct colour-dependent photometric transformation varies from asteroid to asteroid, which increases the photometric uncertainty if the spectral type is unknown. 

Finally, the brightness of an asteroid will change over time due to shape-driven rotational brightness variations, and it may also experience orbital-timescale changes in the observed $H$ due to the `apparition effect'.
This effect can occur for asteroids with non-spherical shapes
as the aspect angle (the angle between the spin axis and the Earth-asteroid vector) changes, meaning the asteroid projects different cross-sectional areas to the observer.
These changes in projected area lead to variations in the absolute magnitude and lightcurve amplitude measured between apparitions \citep{fernandez-valenzuelaModelingLongTermPhotometric2022}.
	However, the influence of rotational and apparition effects can be reduced by deriving $g_\lambda(\alpha)$ from data restricted to a single apparition, sampling the full range of rotational phases. 
	This ensures that any brightness variations are primarily due to rotational effects and results in approximately homogeneous scatter around the true $g_\lambda(\alpha)$.

The current asteroid measurement catalogues held at the Minor Planet Center (MPC) and the JPL Small Bodies Database apply general filter transformations from the reported photometric bandpass to the $V$-band for all large surveys, but they often do not correct for the rotational/apparition effects listed above. 
Given the 
$\sim 4\times10^8$
observations in these databases from different facilities with a range of photometric accuracy and bandpasses, this is not surprising. 
Instead, absolute magnitudes $H_V$ are calculated by transforming to the $V$-band and fitting the parameterised $(H,G)$ system of \citep{bowellApplicationPhotometricModels1989}, where the form of $g_\lambda(\alpha)$ depends only on a phase parameter $G$. 
$H_V$ is calculated for most asteroids using an assumed value of G = 0.15; this approach is used by the MPC and \astorb \citep{moskovitzAstorbDatabaseLowell2022}. 
Often object-specific values of $G$ are only used for the subset of asteroids with highly accurate known phase curves.

The combined uncertainty of photometric accuracy and conversion between bandpasses is important as fitting the phase curve leads directly to the measurement of $H_\lambda$, which in turn allows determination of the diameter $D$ 
\begin{equation}
    D_\mathrm{km}=C_\lambda \frac{10^{-H_\lambda/5}}{\sqrt{p_\lambda}}
\end{equation}
where $p_\lambda$ is geometric albedo and $C_\lambda$ is a wavelength dependent constant related to the absolute magnitude of the Sun; e.g.\ for $V$-band observations $C_V = 664.5\, \si{km}$ \citep{pravecBinaryAsteroidPopulation2007}.
In the past it has been shown that significant biases can exist for $H_\lambda$ \citep{juricComparisonPositionsMagnitudes2002,pravecAbsoluteMagnitudesAsteroids2012}. While this has been presumed to be due to historical photometric calibration or transformation issues, it may also be partly due to ignoring rotational and/or apparition effects on phase curve accuracy.

Aside from absolute magnitude, $g_\lambda(\alpha)$ is also important in the study of asteroid properties.
The form of $g_\lambda(\alpha)$ has been found to be related to both surface composition and asteroid taxonomy \citep{belskayaOppositionEffectAsteroids2000a,oszkiewiczAsteroidTaxonomicSignatures2012a}, and also the surface scattering properties such as regolith particle size and overall roughness of the surface \citep{muinonenAsteroidPhotometricPolarimetric2002}. 
Therefore there are two benefits of measuring $g_\lambda(\alpha)$ beyond establishing $H_\lambda$. 
First, this can indicate approximate compositions for asteroids without multi-colour photometry or spectroscopy. 
Second, as surface composition is related to albedo \citep{usuiALBEDOPROPERTIESMAIN2013}, this can significantly decrease the uncertainty in the estimated diameter of an asteroid when using an assumed albedo.

Recognising the importance of information obtained by measuring the phase curves of asteroids, such measurements of asteroid ensembles have been made by several authors. 
By obtaining high-quality photometry of a small number of objects, \cite{harrisAsteroidLightcurveObservations1989} and \cite{lagerkvistAnalysisAsteroidLightcurves1990} showed that phase function and taxonomy were related. 
Although the observed correlations could be due to either surface composition or albedo as these two parameters are also related, \cite{belskayaOppositionEffectAsteroids2000a} found that albedo was the primary factor influencing the phase curve. 
Albedo is generally the primary factor in determining whether single scattering (for more absorbant dark surfaces) or multiple scattering (for more reflective bright surfaces) of reflected sunlight dominates the observed photometry \citep{muinonenAsteroidPhotometricPolarimetric2002}.
Moving to large scale analysis of much bigger databases of asteroid photometry, \cite{oszkiewiczOnlineMultiparameterPhasecurve2011,oszkiewiczAsteroidTaxonomicSignatures2012a} debiased the existing MPC catalogue for inter-observatory offsets in reported magnitudes, deriving absolute magnitudes and phase functions for approximately $5\times10^5$ asteroids. 
Their work clearly showed correlations between the phase parameters and position in the main belt, similar to those found by colour surveys, although uncertainties caused by the inhomogenity of the source photometry remained. 
\cite{veresAbsoluteMagnitudesSlope2015a} reported phase function fits for over $2\times10^5$ asteroids from photometry obtained using sparse photometry from the wide-field Pan-STARRS survey.
They used a Monte-Carlo technique to estimate parameter uncertainties resulting from the unknown rotational lightcurve modulation. 
Significant uncertainties in the fitted parameters were reported due to this and the sparseness of the dataset.

More recently there have been a number of studies focused on extracting asteroid phase curve parameters from additional wide-field surveys.
In particular \cite{mahlkeAsteroidPhaseCurves2021}, hereafter \citetalias{mahlkeAsteroidPhaseCurves2021}, performed a phase curve analysis for 95,000 asteroids observed by the \ATLAS sky survey \citep{tonryATLASHighcadenceAllsky2018}.
While similar to Pan-STARRS where data is obtained with consistent reduction methods by a single facility (consisting of multiple units), the use of only 2 filters with \ATLAS (compared to 6 with Pan-STARRS), together with a wider field of view per exposure,  allows for a significantly higher cadence of observations.
This makes the \ATLAS dataset an excellent resource for the study of asteroid phase curves, even though its sensitivity is lower than other sky surveys.
Performing Bayesian parameter inference Markov-Chain Monte Carlo analyses of the asteroid phase curves, \citetalias{mahlkeAsteroidPhaseCurves2021} obtained absolute magnitudes and phase curves with robustly estimated uncertainties.
In this paper, we describe an independent phase curve analysis of $9.6\e{7}$ photometric measurements of 4.5\e{5} main belt and Trojan asteroids obtained by the \ATLAS survey, from December 2015 to January 2022. 
By doing so we aim to provide a large dataset of reliable phase curve parameters and absolute magnitudes for the use the asteroid community.
As all observations were obtained using the same facility and filter system, we avoid the potential problems in combining data from multiple sources.
Our work builds upon the recent study by \citetalias{mahlkeAsteroidPhaseCurves2021} as we have access to a longer baseline of observations.
Furthermore we approach the problem in a different manner whereby we consider the phase curves during individual apparitions for all asteroids, rather than fitting a single phase curve to all observations.

In section \ref{sec:methods} we describe the \ATLAS photometric observations and our algorithm to fit phase curves to individual apparitions, including the filtering out of outlying observations and inaccurate phase curves.
In section \ref{sec:results} we present the resulting Selected \ATLAS Phase-curves (SAP) dataset and its distributions of absolute magnitude, phase parameters and colours.
Furthermore we compare our results to those of previous studies.
Section \ref{sec:discussion} demonstrates the power of our dataset by considering the phase curve properties of asteroid families and Jupiter Trojans observed by \ATLAS.
We also discuss the limiting factors in our methodology and the implications for future surveys.
Finally in section \ref{sec:conclusions} we present a summary of the work, highlighting the major results.

\section{Methodology}\label{sec:methods}

\subsection{ATLAS observations}\label{sec:ATLAS_observations}

Until recently, the \ATLAS survey was composed of two $0.5\, \si{m}$ robotic telescopes located on Mauna Loa and Haleakala, Hawaii, with a separation of approximately 140 kilometres. 
Two additional Southern hemisphere stations have recently been added in Chile and South Africa, however the analysis in this work consists exclusively of data obtained with the original Northern hemisphere units. 
The main goal of \ATLAS is to detect potential terrestrial impactors in the Near-Earth Object (NEO) population, including those on their final impact trajectory \citep{tonryATLASHighcadenceAllsky2018}.
It does this by taking images of the night sky on an automated schedule and subtracting a pre-determined background static sky wallpaper, thus leaving behind only photometric sources that differ from the background sky.
Moving objects can be clearly identified in a series of these difference images.
Apart from discovering asteroids, the difference imaging technique has also proved extremely valuable for a range of transient astronomy science; e.g.\ variable stars \citep{heinzeFirstCatalogVariable2018}, supernovae \citep{smithDesignOperationATLAS2020}, gravitational wave source identification \citep{smarttKilonovaElectromagneticCounterpart2017} and also very slow motion distant Solar System bodies \citep{dobsonPhaseCurvesKuiper2023}.

Each \ATLAS unit consists of a custom Wright-Schmidt design with $0.5\, \si{m}$ aperture and $0.65\, \si{m}$ spherical primary mirror \citep{tonryATLASHighcadenceAllsky2018}.
The detector is an STA-1600 $10.5 \times 10.5\, \si{k}$ CCD with a pixel scale of $1.86\arcsec$.
\ATLAS primarily surveys in an ``orange'' $o$ and ``cyan'' $c$-filter, reaching down to $\sim 19.5\, \si{mag}$ for its typical $30\, \si{s}$ exposures.
These filters are approximately equivalent to PS1 filters $r+i$ ($560-820\, \si{nm}$) and $g+r$ ($420-650\, \si{nm}$) respectively.
Generally the bluer $c$-filter is reserved for new moon dark time during survey operations, whereas the redder $o$-filter is more frequently used during bright time observations \citep{smithDesignOperationATLAS2020},
although this strategy has evolved over the course of the survey.

\ATLAS units scan across the available night sky (with an accessible declination range of $-45^{\circ} < \delta < +90^{\circ}$ for the Hawaii units), revisiting each field 4 times over timescales $\leq 1$ hr.
An individual unit can cover approximately 1/4 of the visible sky per night in this manner, therefore the two Hawaii units together cover the accessible Northern sky every 2 days.
The addition of units in South Africa and Chile opens up the Southern sky to \ATLAS, increasing all-sky coverage and cadence of field revisits and allows \ATLAS to approach $24\, \si{hr}$ night sky monitoring for Potentially Hazardous Asteroids (PHAs)
and reducing restrictions on observations due to local weather.
Telescope operations at each unit are fully automated (for monitoring weather, dome control and scheduling) and each unit will proceed with observations when safe and able to do so.

The \ATLAS pipeline automatically reduces images and performs sky subtraction to search for transient sources via difference imaging. 
Photometry of detected sources is measured from the original reduced and calibrated image \citep[using \dophot,][]{schechterDoPHOTCCDPhotometry1993,alonso-garciaUNCLOAKINGGLOBULARCLUSTERS2012}.
Moving object detection and linking of these sources is performed using the Moving Object Processing System \citep[MOPS,][]{denneauPanSTARRSMovingObject2013}.
Candidate moving object tracklets, consisting of $>3$ (typically 4) detections, are matched against known MPC sources (via \MPChecker\footnote{\url{https://www.minorplanetcenter.net/cgi-bin/checkmp.cgi}}).
Candidate NEOs are assigned an NEO digest metric score \citep{keysDigest2NEOClassification2019} and are filtered by a deep learning classifier \citep{chybarabeendranTwostageDeepLearning2021} prior to screening by a human observer and submission to the MPC.

\ATLAS acts in synergy with the other major sky surveys, such as Pan-STARRS, 
enhancing sky coverage for NEO discovery but also obtaining serendipitous observations of any minor planets in the survey footprint.
In addition to prompt NEO astrometry, \ATLAS tracklets that have been matched against the ephemerides of known objects are  submitted en masse to the MPC at the end of every night. 
As such, in addition to its primary mission of NEO discovery and characterisation, \ATLAS is a valuable source of astrometry and photometry for a significant number of Solar System objects.

\subsection{The \rockAtlas database} \label{sec:rockAtlas}

For this study we generated a database of \ATLAS detections matched to known Solar System minor planets.
The \rockAtlas database \citep{Young_rockAtlas} has been processing all \ATLAS observations for a fixed list of known asteroids since 2017.
The initial list of asteroids to be tracked was $\sim445,000$ asteroids from the \astorb\footnote{\url{https://asteroid.lowell.edu/main/astorb/}} dataset \citep{moskovitzAstorbDatabaseLowell2022} with well defined orbits at that time.
Accurate orbital parameters were required as \rockAtlas calculates the ephemerides of these objects and matches them to \ATLAS observations.
When \rockAtlas is running, a night's worth of observations is downloaded from the \ATLAS automated detection and photometry pipeline - \dophot \citep{tonryATLASHighcadenceAllsky2018}.
\dophot provides details on all \ATLAS detections that night, such as time, position, magnitude etc.
These detections consist of moving Solar System bodies, static transients and false detections.
\rockAtlas calculates precise ephemerides for all asteroids in its watch list at the mid-time of each exposure using \pyephem\footnote{\url{https://rhodesmill.org/pyephem/}} and \orbfit\footnote{\url{http://adams.dm.unipi.it/orbfit/}}.
These ephemerides are then cross-matched to the \dophot detection positions using a \healpix grid \citep{gorskiHEALPixFrameworkHigh2005,zoncaHealpyEqualArea2019}. 
A \dophot detection is associated with a moving object, provided the separation from the predicted ephemeris at the exposure mid-time is less than $5\, \si{arcsec}$, and it is added to the \rockAtlas database.
Together with the ephemerides, \pyephem and \orbfit provide the viewing geometry for each observation including heliocentric and geocentric distances, the Sun-asteroid-Earth phase angle and galactic latitude, which are appended to the database.

The \rockAtlas database provides an ever growing dataset of \ATLAS asteroid photometry with all the additional information needed for studying asteroid phase curves. It contains $\sim 9.6 \e{7}$ observations as of January 2022, comprised of $\sim65\%$ in the $o$-filter and $\sim35\%$ in the $c$-filter.
Asteroids in the \rockAtlas database have a median of $\sim130$ observations.
Apparent magnitudes range from $\gtrsim 12\, \si{mag}$ down to a limiting magnitude of $\sim 19.5$, with a median observed magnitude of $\sim 17\, \si{mag}$.
Overall the median photometric uncertainty is $0.05$ magnitudes, 
however this uncertainty increases substantially at fainter magnitudes;
the median photometric uncertainty is 0.1 magnitudes for $m_o >17\, \si{mag}$. 

\begin{figure}
    \includegraphics{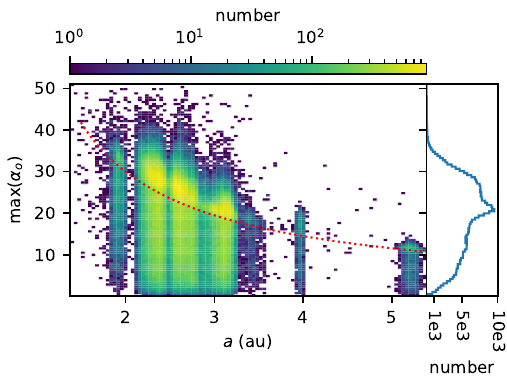}
    \caption{
    Two dimensional histogram showing the number distribution of the maximum observed phase angle of asteroids in the \rockAtlas database ($o$-filter) as a function of the object's semimajor axis (left-hand panel).
    The maximum phase angle for a zero-inclination circular orbit, $\text{max}(\alpha) = \sin^{-1}{1/a}$, is shown as a dotted line.
    The total distribution of maximum phase angle for the whole dataset is shown as a vertical histogram in the right-hand panel.
    The Main Asteroid Belt dominates the database; these asteroids are typically observed up to a maximum phase angle of $\gtrsim 20^\circ$.
    }
    \label{fig:phase_angle_max}
\end{figure}

In this work we focus on the phase curve properties of asteroids from the main asteroid belt out to the Jupiter Trojans, which dominate the \rockAtlas dataset (figure \ref{fig:phase_angle_max}). 
Other studies have been performed of Near Earth Objects ($q<1.3\, \si{AU}$) and Outer Solar System Objects ($a>Q_\text{Jup} = 5.457 \, \si{AU}$) observed with \ATLAS \citep{heinzeNEOPopulationVelocity2021, dobsonNewIncreasedCometary2021,dobsonPhaseCurvesKuiper2023}.
The median maximum phase angle of observations is $\sim 20^\circ$, which is the expected value for asteroids on circular orbits within the outer main belt.
As expected and shown in figure \ref{fig:phase_angle_max} the maximum observed phase angle decreases with semimajor axis.
The more numerous Main Belt asteroids typically have observations up to a phase angle of $\sim20^\circ$ whereas the most distant objects we consider in this work, the Jupiter Trojans, are observed only up to $\alpha \sim 10^\circ$.
Observations of asteroid brightness at low phase angles ($<5^\circ$) are essential to characterise any opposition brightness surge that may be present in the phase curve as the phase angle decreases to zero.
The median minimum observed phase angle for objects in the \rockAtlas database is $\alpha \sim 3^\circ$, making this a valuable dataset for accurately investigating asteroid phase curves.
One may refer to table \ref{tab:data_overview} for details of the observational circumstances of a sample of objects in the database.

Recently, direct public access to \ATLAS asteroid photometry has been made available via the online SSCAT database\footnote{\url{https://astroportal.ifa.hawaii.edu/atlas/sscat/}}.
This database is constructed in a similar manner to \rockAtlas (by matching \dophot photometry to asteroid ephemerides) and therefore it is recommended that the same considerations addressed in this work (e.g.\ dealing with outlying photometry, see section \ref{sec:phase_curve_fits}) should be applied to future work making use of the SSCAT database.

Alternatively, the \ATLAS forced photometry server 
\citep{shinglesReleaseATLASForced2021} provides access to \ATLAS observations for static and moving astrophysical objects.
However, photometry retrieved by forced measurement at requested ephemerides \citep[such as the method used by][]{dobsonPhaseCurvesKuiper2023} may differ from the automatically generated \dophot measurements. 
In addition, the computational overhead of performing forced photometry queries would most likely limit the number of objects we wish to consider, therefore we did not make use of this tool in this work.
For example, a forced photometry query for a typical MBA takes approximately 20 minutes to match frames and determine magnitudes, whereas the same query through \rockAtlas takes only seconds to retrieve magnitudes, distances and phase angles as many of the parameters have already been calculated and stored in the database. 

\subsection{Phase curve functions} \label{sec:phase_curve_functions}

There have been many phase curve models devised in differing degrees of complexity and physical basis. For this work we concentrate on two models. The first is the $(H,G)$ system published by \cite{bowellApplicationPhotometricModels1989}, hereafter denoted as \citetalias{bowellApplicationPhotometricModels1989}. This model has two primary advantages. It is based on the observed surface scattering properties of well observed asteroids. 
Also, having been in use for over 3 decades, it has acted as the {\it de-facto} standard representation of asteroid photometric properties over that period. 
Phase curves are fitted to the reduced magnitude, $V(\alpha)$, which is the apparent magnitude corrected for distance effects such that only the phase function $g_\lambda(\alpha)$ from equation \ref{eqn:absolute_magnitude} leads to deviation from the absolute magnitude
\begin{equation}
V(\alpha) = m - 5 \log(R\Delta)
\label{eqn:reduced_magnitude}
\end{equation}
In the \citetalias{bowellApplicationPhotometricModels1989} system the change in reduced magnitude with respect to phase angle is described by
\begin{equation}
    V(\alpha)=H-2.5\log\left( \left[1-G\right]\Phi_{\text{B89},1}(\alpha) + G\Phi_{\text{B89},2}(\alpha)\right) \label{eqn:B89_model}
\end{equation}
where $H$ is the absolute magnitude of the object (brightness at zero phase scaled to a distance of $1\, \si{au}$), $\Phi_{\text{B89},i}$ are the basis functions, and $G$ is the phase parameter that controls the form of the model phase curve.

The second model we use is the $(H,G_{12}^*)$ system from \cite{penttilaG1G2Photometric2016}, hereafter referred to as \citetalias{penttilaG1G2Photometric2016}. 
This was derived from the new IAU standard $(H,G_1, G_2)$ system 
formulated by \cite{muinonenThreeparameterMagnitudePhase2010a}, which more accurately reflects the phase function at small phase angles and is defined as
\begin{equation}
\begin{split}
        V(\alpha) = & H-2.5\log(G_1\Phi_{\text{P16},1}(\alpha) + G_2\Phi_{\text{P16},2}(\alpha) \\
        & + \left[1-G_1-G_2\right]\Phi_{\text{P16},3}(\alpha)) \label{eqn:P16_model}
\end{split}
\end{equation}
where $\Phi_{\text{P16},i}$ are the basis functions in this system.
In particular $\Phi_{\text{P16},3}$ models the opposition effect.
There are two phase parameters $G_1$, $G_2$ and in the \cite{penttilaG1G2Photometric2016} model these parameters are combined via the single $G_{12}^*$ phase parameter; such that
$G_1 = 0.84293649 G_{12}^*$ and $G_2 = 0.53513350(1-G_{12}^*)$.
The \citetalias{penttilaG1G2Photometric2016} model is better suited for modelling the phase curves of objects with lower quality, more sparse observations.
The phase parameter $G_{12}^*$ must not be confused with the similar $G_{12}$ parameter which was also defined by \cite{muinonenThreeparameterMagnitudePhase2010a}.
In this work where we consider mainly the \citetalias{penttilaG1G2Photometric2016} model, we use the notation $G_{12}^* \rightarrow G_{12}$ for simplicity.
Likewise, from here on we use $H_{12}$ to distinguish the \citetalias{penttilaG1G2Photometric2016} absolute magnitude from the \citetalias{bowellApplicationPhotometricModels1989} absolute magnitude $H$.

\subsection{Phase curve fitting algorithm} \label{sec:phase_curve_fits}

Using the photometric observations and associated viewing geometries in the \rockAtlas database we have measured the phase curve behaviour of asteroids in both the $o$ and $c$ filters.
For a given asteroid, all associated \ATLAS observations, ephemerides and related quantities are retrieved from \rockAtlas.
We used the \sbpy package \citep{mommertSbpyPythonModule2019} to fit various phase curve models to these observations, namely $HG$ \citepalias{bowellApplicationPhotometricModels1989}
and $H_{12}G_{12}$ \citepalias{penttilaG1G2Photometric2016}.
\\

We build upon previous studies of large numbers of asteroid phase curves by accounting for the apparition effect in asteroid brightness.
We define an apparition as the period of time when an asteroid is continuously visible in the sky to the \ATLAS survey, i.e. when it is not too close to the sun for observation.
We therefore group observations according to their solar elongation angle and consider each apparition of an asteroid separately when fitting phase curves \citep[similar to][hereafter \citetalias{schemelZwickyTransientFacility2021}]{schemelZwickyTransientFacility2021}.
The long baseline of the \ATLAS survey from 2017 to the present day means many asteroids have been observed for up to 6 different apparitions and there are asteroids observed by \ATLAS which clearly exhibit the apparition effect as previously noted by \citetalias{mahlkeAsteroidPhaseCurves2021}.
Furthermore, the presence of such an effect implies the asteroid has a non-spherical shape and a tilted spin pole and so we would expect to observe measureable lightcurve variations.
We can see the change in lightcurve amplitude with apparition as the projected cross-sectional area changes \citep{fernandez-valenzuelaModelingLongTermPhotometric2022}, e.g.\ for asteroid (766) Moguntia in figure \ref{fig:apparitions_Moguntia}.
Identifying such objects for further study with targeted lightcurve observations and more sophisticated lightcurve inversion methods would be extremely useful to determine accurate physical properties.
For example, work by \cite{durechAsteroidModelsReconstructed2020} has already demonstrated the benefits of combining sparse \ATLAS data (which covers a range of different viewing geometries to help with shape determination) with dense lightcurve data (which helps constrain the asteroid rotation) using convex lightcurve inversion (CLI) methods. 
\\

Due to the automated nature of \ATLAS \dophot processing and the matching of observations by \rockAtlas, anomalous magnitude measurements 
exist in the database.
Such outlying data points can be caused by false matching of non-asteroid sources, background star contamination, cosmic rays, stellar diffraction spikes and ghost reflections from bright sources etc.\ \citep{heinzeNEOPopulationVelocity2021}.
In the case of false matching, \dophot photometric detections may have been erroneously matched to an asteroid by the \rockAtlas code, either due to inaccurate ephemerides or the matching of a nearby stellar/galactic source when the asteroid is fainter than predicted and is not detected (e.g.\ at its rotational lightcurve minimum).
Therefore the first step was to reject any data points where the separation of the \orbfit ephemerides to the \dophot position was $\theta>1\, \si{arcsec}$ to reduce the number of false matches. 
In addition we rejected any observations with low galactic latitude, $b\leq10^\circ$, as near the galactic plane the higher density of background stars may either contaminate the measured brightness or lead to \rockAtlas-\dophot matching errors. We also rejected any photometry that was performed on frames with a recorded 5-sigma limiting magnitude of $<17.5$ (e.g.\ observations made near full moon).
Finally we rejected data points with extremely low (or zero) uncertainties in the \dophot magnitude, $\sigma_m\leq 0.005\, \si{mag}$, as we do not trust such measurements given the typical photometric accuracy of \ATLAS.
Out of this series of filters, the ephemeris position check removed the bulk of obviously erroneous observations (25\% of all detections had $\theta > 1$ arcsec), therefore this tighter constraint is worth possibly rejecting the small number of true matches with separations $1<\theta(\si{arcsec})<5$.

We then performed additional filtering of outlying data to each individual apparition, considering the $o$ and $c$-filters separately.
To each apparition we fitted the \citetalias{bowellApplicationPhotometricModels1989}
$HG$ phase curve model using a Levenberg-Marquardt non-linear least squares algorithm \citep[\astropy;][]{astropycollaborationAstropyProjectSustaining2022}.
For this first stage of phase curve fitting the data points were not weighted by their photometric uncertainties, in order to reduce the influence of extreme outliers in brightness with very low uncertainties.
The initial estimate for $H$ was taken from \astorb and colour corrected from $V$-band to the \ATLAS filter of the observations.
This conversion depends on an asteroid's colour and taxonomic class but we use the values from \cite{heinzeNEOPopulationVelocity2021} which are derived from the mean colour of S and C type asteroids.
\begin{align}
	\langle V - o \rangle & = 0.332 \\
	\langle V - c \rangle & = 0.054
\end{align}
The slope parameter $G$ was set at $G=0.15$ which is the standard assumed value for all asteroids \citep{marsdenNotesIAUGeneral1985}.
Using this $HG$ fit we rejected any observations with a residual greater than 3 magnitudes to remove the most extreme outliers.
This limit is higher than the maximum lightcurve amplitude for asteroids in the Asteroid Lightcurve Database \citep[LCDB;][]{warnerAsteroidLightcurveDatabase2021} and hence only rejects photometry that is unrealistic for rotating inert asteroids.
We note that
\citetalias{mahlkeAsteroidPhaseCurves2021} used a residual limit of 1 magnitude when filtering their \ATLAS data.
We then calculated the standard deviation of the remaining residuals and performed a 2-sigma clip to remove further outliers.
By considering apparitions separately and fitting each for $H$ we account for possible apparition effects which would increase the residuals if all observations were considered together.
If an asteroid displays large shifts in $H$ between apparitions a phase curve fit to the full set of observations would not reflect the true residuals at each apparition, therefore outlier rejection could erroneously remove part or all of a particular apparition.
\\

After cleaning up the observations for an asteroid we consider the properties of each apparition and selected a ``primary'' epoch to use for the main phase curve fit.
We scored each apparition by the total number of observations, the number of observations at phase angles $\alpha <5^\circ$ and by the range of phase angle coverage.
The scores for each of these parameters were scaled by subtracting the mean and scaling to unit variance.
The total score of an apparition was taken to be the sum of all three scores (with equal weighting).
The primary apparition was the one which had the highest total score and so the best combination of all three constraints required to get a reliable phase curve fit.
Simply selecting the apparition with the most observations does not guarantee good coverage of the opposition effect at low phase angles and will not accurately constrain the slope parameter and absolute magnitude.
We only selected primary apparitions in the $o$-filter as the \ATLAS dataset preferentially contains observations in $o$ rather than $c$.

The primary apparition was then fitted for absolute magnitude and phase curve parameters with the selected phase curve model from \sbpy, where the initial values for $H$ are taken from \astorb and colour corrected (as described in the data cuts above).
Our code fits all available \sbpy models but we restrict our analysis in this work to \citetalias{bowellApplicationPhotometricModels1989} and \citetalias{penttilaG1G2Photometric2016} for simplicity.
The Levenberg-Marquardt least squares algorithm was used to fit the phase curve model to the data, now weighting each data point by its photometric uncertainty.
The uncertainties in the fitted phase curve parameters are taken to be the square root of the diagonal of the covariance matrix.
All other apparitions in $o$ and $c$ are fitted for $H$ only, with the slope parameters kept fixed at the value obtained from the primary apparition fit.
Therefore we obtain the phase curve slope parameter from the primary apparition alone and implicitly assume that the phase curve is the same in $o$ and $c$ within our measurement uncertainties.
In our investigation we originally attempted full phase curve fits (absolute magnitude and slope parameter) to all observations, and also to each apparition.
For most objects this can be a valid approach as oftentimes outlying photometry is easily caught with the steps above and there is little variation between apparitions.
However, for some objects missed outliers and apparitions with poor observational coverage can skew the overall phase curve fit.
We found that obtaining the slope parameter from the primary apparition alone did not lead to significant changes for most objects, however, this approach did succeed in removing the smaller number of objects with clearly unphysical slope parameters.

Unlike previous studies we do not constrain slope parameters when fitting the \citetalias{bowellApplicationPhotometricModels1989} and \citetalias{penttilaG1G2Photometric2016} models.
Normally the slope parameters are restricted to ``physical'' values such that the phase curve is always monotonically increasing in brightness as phase angle decreases.
Imposing such constraints can lead to a distribution in phase parameter with pile ups near the boundaries \citepalias[e.g. at 0 and 1 for $G_{12}$,][see their figure 5]{mahlkeAsteroidPhaseCurves2021}.
These edge cases may tend towards the boundaries because they genuinely have surfaces described by extremely low/high values of phase parameter.
Otherwise they may have insufficient observational coverage to accurately constrain the phase parameter, or very extreme rotational variations affecting the phase curve fit.
Significant changes in viewing geometry and/or observed aspect of an asteroid can also lead to deviations from expected phase curve behaviour even after correction for rotational effects \citep{jacksonEffectAspectChanges2022a}.
Therefore to catch these possible interesting cases and to prevent an artificial build up of phase parameter at the boundaries we allow the fitter to go to these ``unphysical'' values.
\\

Importantly, we do not consider rotational variations in brightness in our phase curve model, similar to many previous studies of asteroid phase curves \citep[e.g.][]{mcneillComparisonPhysicalProperties2021,mahlkeAsteroidPhaseCurves2021}. 
We assume that the variation in reduced magnitude due to lightcurve amplitude will be reflected in the formal uncertainties of the phase curve fit and in most cases this amplitude is low compared to variation caused by photometric uncertainty.
The cadence of \ATLAS data alone can in some cases be used to find unique rotation periods.
For example \cite{durechAsteroidModelsReconstructed2020} were able to determine rotation periods from \ATLAS data alone for $\sim5000$ asteroids out of an initial sample of $\sim180,000$ objects.
Furthermore \cite{erasmusDiscoverySuperslowRotating2021} searched \ATLAS observations of $\sim250,000$ asteroids for extreme slow rotators ($\mathrm{period} > 1000\ \si{hours}$) and identified 39 objects with high-confidence period solutions.
However, additional dense in time photometry can assist greatly in determining unique rotation states and rejecting alias effects, as shown by \cite{durechAsteroidModelsReconstructed2020} who used periods from the LCDB to significantly reduce the parameter space to be searched when determining the shape and spin state from \ATLAS data.

\subsection{Filtering of phase curves} \label{sec:phase_curve_classification}

We implemented our phase curve fitting algorithm in a relatively agnostic manner to the observational data, which has varying coverage and quality.
As such, we know that not all fits in our full phase curve database will be reliable and we need to identify the high quality fits with well defined parameters.
To aid this, we recorded a variety of metrics for each phase curve fit, including the number of data points in each fitted apparition, the phase angle coverage and the statistics of the residuals for the fitted phase curve model.
We used the values of these metrics to select and reject phase curves and we tuned the acceptable values for each metric by considering the final distributions of $G$, and its uncertainty $\sigma_G$ for the \citetalias{bowellApplicationPhotometricModels1989} fits.
Visual inspection of individual phase curves showed that the most extreme (and probably unphysical) values of $G$ and $\sigma_G$ were caused by poor phase angle coverage or large photometric variations which meant the observations could not accurately constrain the phase curve behaviour of an asteroid.
The final acceptable values for each metric were therefore chosen such that the vast majority of objects had $0<G<1$, without us having to perform a hard cut on $G$ at these boundaries.
The details of this process are given in appendix \ref{appendix:quality_control_metrics} but the main restrictions were such that the primary apparition had to have at least 50 observations, with at least one observation at low phase angle, $\alpha <5^\circ$, after outlier rejection.
All other apparitions required at least 10 observations to be considered reliable.
After developing  our metric cuts for the $HG$ model, we then applied them in the same manner to the $H_{12}G_{12}$ data. Our intention was to not introduce biases by over-tuning the metric parameters for a particular model, and to allow for a less biased comparison between the two models.

By ensuring sensible distributions of phase curve parameters and uncertainties for the overall dataset, when we take a large subsample for a particular dynamical class we can assume the statistics of that distribution will be a good representation of the overall properties of the class.
Poorly constrained phase curve parameters for individual objects may still exist, but with large enough numbers we can compare the overall trends for different populations in a statistical manner.

\subsection{Apparition effects, absolute magnitudes and asteroid colours in ATLAS filters} \label{sec:app_abs-mag_colours}

\begin{figure*}
    \includegraphics{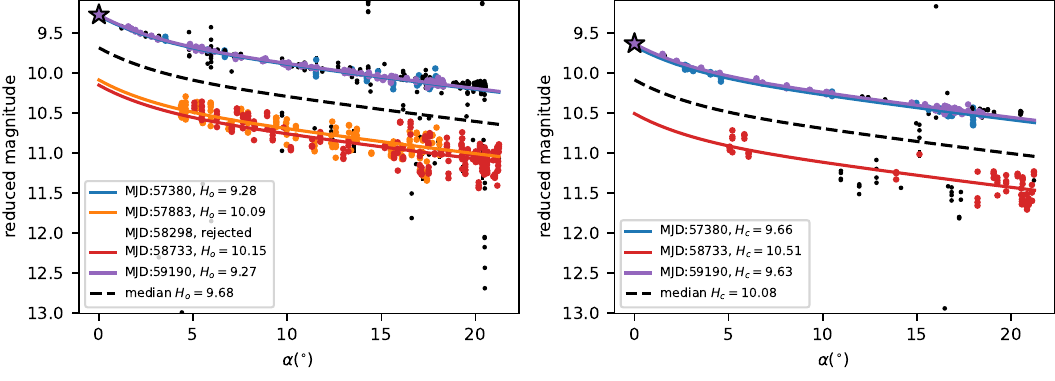}
    \caption{
    \ATLAS phase curves of asteroid (766) Moguntia, an object displaying a large apparition effect.
    Observations in the $o$-filter are shown in the left-hand panel. 
    Each apparition has been colour coded and outlying photometry/rejected apparitions are shown as small markers (some of which are obscured by the figure legend).
    The limits of the $y$-axis have been set to focus on retained photometry, ignoring any more extreme outliers.
    The corresponding $HG$ phase curve fit for each apparition is shown by the matching coloured curves, with the primary apparition indicated by the coloured star.
    The median $H$ of the retained apparitions is indicated by the dashed curve with parameters: $H_o=9.68$, $G=0.21$.
    Observations in the $c$-filter are shown in the right-hand panel, where marker/line colour have the same meanings as the left panel.
        The large shifts in $H$ between apparitions are clear ($F_\text{app}=5.9$; equation \ref{eqn:app_check}).
    }
    \label{fig:apparitions_Moguntia}
\end{figure*}

By considering phase curve fits of individual apparitions, we are able to identify asteroids with significant variations in absolute magnitude between apparitions.
In figure \ref{fig:apparitions_Moguntia} we show an example of the apparition effect for asteroid (766) Moguntia. 
The presence of this effect in \ATLAS observations was noted by \citetalias{mahlkeAsteroidPhaseCurves2021}, and additionally by \citetalias{schemelZwickyTransientFacility2021} who also fitted separate apparitions in their study of the Jupiter Trojans using Zwicky Transient Facility \citep[ZTF,][]{bellmZwickyTransientFacility2019} data.
For each asteroid we used the statistics of the measured $H$ and the phase curve residuals to identify significant variations across several apparitions.
We define a parameter
\begin{equation}
    F_\textrm{app}=\frac{\sigma(H)}{{\rm median}\left[\sigma(m_\textrm{obs}-m_\textrm{fit})\right]} 
    \label{eqn:app_check}
\end{equation}
which is the ratio of the standard deviation of $H$ across the retained apparitions, divided by the median of the standard deviation of the observed minus fitted phase curve residuals in each apparition.
This ratio compares the variation of $H$ across several apparitions to the typical brightness variation after phase curve correction, which is dominated by rotational lightcurve effects and photometric uncertainty.
We consider just the $o$-filter data in this test due to the larger number of photometric measurements compared to the $c$-filter.
When the changes in $H$ are larger than the phase curve corrected brightness variations, we assume that the object is undergoing apparition effects.
By inspecting the phase curves of individual objects we found that a boundary of $F_\textrm{app}\geq0.9$ succeeds in identifying the objects with the strongest apparition effects.
The remaining asteroids with $F_\textrm{app}<0.9$ do not present a significant apparition effects compared to the rotational variation and \ATLAS photometric accuracy.
Also, we note that 28\% of the asteroids have only one suitable apparition after the quality cuts and so cannot be searched for apparition effects.
Using the $F_\textrm{app}\geq 0.9$ metric,  we found that 9\% of our asteroids displayed a significant apparition effect, where this is clearly a lower limit to the true number that exhibit this phenomenon.

\begin{figure*}
    \includegraphics{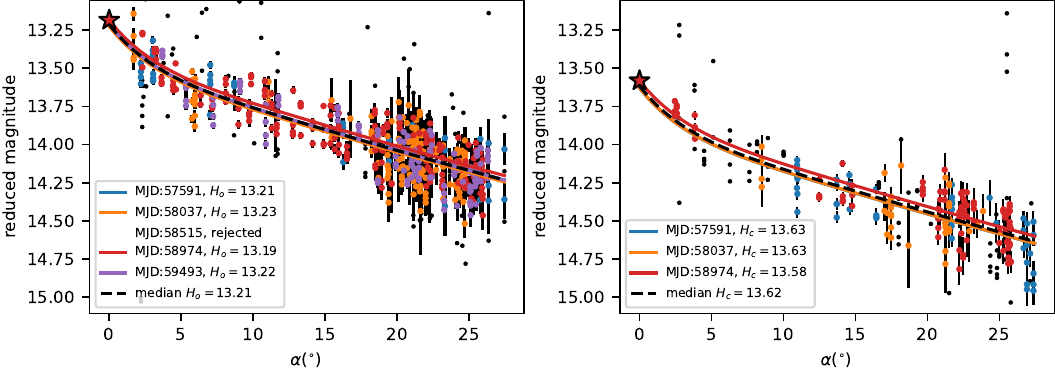}
    \caption{
    Phase curves for (4986) Osipovia, a typical MBA observed by \ATLAS.
       As in figure \ref{fig:apparitions_Moguntia} the left and right-hand panels show the $o$ and $c$-filter observations respectively.
    Marker/line colour indicates the apparition to which each observation belongs, with the primary apparition marker by the coloured star and the median value of $H$ indicated by the dashed curve with parameters: $H_o=13.21$, $G=0.30$
    }
    \label{fig:apparitions_Osipovia}
\end{figure*}

For objects without strong apparition effects the final value of absolute magnitude is estimated as the median $H$ of all retained apparitions in either the $o$ or $c$-filters; figure \ref{fig:apparitions_Osipovia} shows this for typical MBA (4986) Osipovia.
We used the median to reduce the influence of possible outlying apparitions that might skew the mean.
For objects with strong apparition effects we should consider the minimum value of $H$, i.e.\ the brightest estimate is closest to the true size.
However this value is only an upper limit on the true absolute magnitude (i.e.\ a lower limit on equivalent size) and such objects require further investigation to accurately determine their physical parameters.
As such we generally exclude the objects with strong apparition effects ($F_\text{app}\geq0.9$) from our results when discussing absolute magnitude and colour, although we make use of the phase parameters $G$, $G_{12}$ of these objects when $H$ is not considered.
For all objects in our dataset the phase parameter is determined solely from the phase curve fit of the primary apparition and so can be compared regardless of apparition effects.

Most of our analysis thus far has focused on the more frequently made observations in the \ATLAS $o$-filter.
The majority of asteroids observed by \ATLAS also have a smaller number of measurements in the bluer $c$-filter.
Thus we can compare the estimates of $H$ in each filter to estimate a $H_c-H_o$ colour for a large sample of asteroids.
Measurements in the $c$-filter suffer from the same systematic errors discussed above for the $o$-filter.
Therefore we calculated colour only for objects with apparitions that satisfy the quality control requirements in both $o$ and $c$-filters, excluding any objects displaying strong apparition effects in the $o$-filter ($F_\textrm{app}\geq 0.9$) as we cannot define a unique $H$ for these objects.
The colour was then calculated as median($H_c$) - median($H_o$).
The \ATLAS colour provides an additional diagnostic for comparing the surface properties of different asteroid populations.

\section{RESULTS} \label{sec:results}

\subsection{ATLAS phase curve database} \label{sec:phase_curve_database}
    
\begin{table*}  
\begin{tabular}{lrrrrrrrrrrrr}
\toprule
    Name & Number & $N_o$ & $N_c$ &  max($t$) (MJD) & $N_\textrm{app}$ & max($\alpha_o$) & max($\alpha_c$) & $\Delta\alpha_o$ & $\Delta\alpha_c$ & $a$ (au) &  $e$ &  $i (^{\circ})$ \\
\midrule
Moguntia &    766 &  1029 &   363 &    59432.262037 &                5 &           21.24 &           21.22 &            19.68 &            19.73 &     3.02 & 0.10 &       10.090543 \\
 Priamus &    884 &  1297 &   420 &    59523.217303 &                6 &           12.60 &           12.59 &            11.53 &            12.26 &     5.20 & 0.13 &        8.906102 \\
Marietta &   2144 &  1309 &   433 &    59600.229942 &                5 &           21.55 &           21.36 &            21.36 &            20.44 &     2.87 & 0.06 &        2.833143 \\
 Yatskiv &   2728 &  1192 &   335 &    59600.242836 &                5 &           29.29 &           29.21 &            28.32 &            28.02 &     2.46 & 0.17 &        2.610173 \\
Osipovia &   4986 &   652 &   246 &    59596.522535 &                5 &           27.46 &           27.37 &            25.76 &            24.92 &     2.56 & 0.21 &        5.611378 \\
 1991 EN &   9857 &  1065 &   314 &    59598.310255 &                6 &           11.57 &           11.45 &            10.68 &             9.28 &     5.16 & 0.03 &       19.612052 \\
\bottomrule
\end{tabular}

\caption{Overview of the observations made by \ATLAS and collected by \rockAtlas, for a selected sample of 6 asteroids 
			(this table is available in its entirety online as supplementary data and \href{https://doi.org/10.17034/db29dc08-5e21-403f-93f3-d60073b97916}{at this link}).
We provide the name and MPC number of each asteroid, alongside the number of detections in the $o$ and $c$-filters ($N_o$, $N_c$) and the date of the last observation used in this work, $\text{max}(t)$.
The maximum phase angle observed in each filter is given by $\text{max}(\alpha_o)$ and $\text{max}(\alpha_c)$, and the maximum minus minimum phase angle range for each filter is $\Delta\alpha_o$, $\Delta\alpha_c$.
Finally, the orbital elements obtained from \astorb are also included; semimajor axis $a$, eccentricity $e$ and inclination $i$.
}
\label{tab:data_overview} 
\end{table*}

\begin{table*}  
\tiny
\begin{tabular}{p{0.7cm}p{0.3cm}p{0.3cm}p{1.0cm}p{0.8cm}p{0.8cm}p{0.3cm}p{0.3cm}p{0.3cm}p{0.3cm}p{0.7cm}p{0.7cm}p{0.3cm}p{0.7cm}p{1.4cm}p{0.3cm}p{0.3cm}}
\toprule
    Name & Number & $n_\textrm{app}$ &  $t_{\textrm{app},0}$ (MJD) & max($\alpha_{\textrm{app},o}$) & max($\alpha_{\textrm{app},c}$) & $N_{\textrm{fit},o}$ & $N_{\textrm{fit},c}$ & $H_o$ & $H_c$ & $\sigma_{Ho}$ & $\sigma_{Hc}$ &  $G$ & $\sigma_G$ & Primary Apparition & Keep$_o$ & Keep$_{c-o}$ \\
\midrule
Moguntia &    766 &                0 &                57380.664873 &                          17.90 &                          18.00 &                   47 &                   59 &  9.28 &  9.66 &      6.93e-03 &      3.65e-03 & 0.21 &        NaN &                  0 &        1 &            1 \\
Moguntia &    766 &                1 &                57883.625009 &                          18.09 &                          17.34 &                  146 &                   20 & 10.09 & 10.67 &      8.86e-03 &      2.34e-02 & 0.21 &        NaN &                  0 &        1 &            0 \\
Moguntia &    766 &                2 &                58298.591274 &                          20.52 &                          20.53 &                  199 &                   49 &  9.25 &  9.61 &      2.00e-03 &      4.54e-03 & 0.21 &        NaN &                  0 &        0 &            0 \\
Moguntia &    766 &                3 &                58733.610938 &                          21.24 &                          21.22 &                  232 &                   70 & 10.15 & 10.51 &      8.96e-03 &      1.61e-02 & 0.21 &        NaN &                  0 &        1 &            1 \\
Moguntia &    766 &                4 &                59190.638252 &                          18.29 &                          18.46 &                  244 &                  138 &  9.27 &  9.63 &      2.42e-03 &      1.22e-03 & 0.21 &   2.87e-03 &                  1 &        1 &            1 \\
 Priamus &    884 &                0 &                57384.434340 &                           9.92 &                           9.94 &                   41 &                   74 &  8.59 &  8.99 &      1.93e-02 &      1.54e-02 & 0.23 &        NaN &                  0 &        0 &            0 \\
 Priamus &    884 &                1 &                57673.634360 &                          10.30 &                          10.30 &                  145 &                   88 &  8.43 &  8.86 &      5.70e-03 &      1.13e-02 & 0.23 &        NaN &                  0 &        1 &            1 \\
 Priamus &    884 &                2 &                58037.632049 &                          10.94 &                          10.78 &                  179 &                   22 &  8.39 &  8.72 &      3.90e-03 &      1.14e-02 & 0.23 &        NaN &                  0 &        0 &            0 \\
 Priamus &    884 &                3 &                58460.627613 &                          11.82 &                          11.82 &                  292 &                   35 &  8.39 &  8.75 &      9.36e-03 &      9.19e-03 & 0.23 &   1.48e-02 &                  1 &        1 &            1 \\
 Priamus &    884 &                4 &                58867.638704 &                          12.60 &                          12.59 &                  200 &                   80 &  8.50 &  8.86 &      7.11e-03 &      1.27e-02 & 0.23 &        NaN &                  0 &        1 &            1 \\
 Priamus &    884 &                5 &                59255.640729 &                          12.05 &                          12.37 &                   20 &                   14 &  8.55 &  8.96 &      2.54e-02 &      1.50e-02 & 0.23 &        NaN &                  0 &        0 &            0 \\
Marietta &   2144 &                0 &                57481.601935 &                          20.26 &                          20.39 &                   45 &                  112 & 10.99 & 11.39 &      2.12e-02 &      1.33e-02 & 0.21 &        NaN &                  0 &        1 &            1 \\
Marietta &   2144 &                1 &                57943.566586 &                          21.55 &                          21.36 &                  175 &                   27 & 11.06 & 11.39 &      1.14e-02 &      2.48e-02 & 0.21 &        NaN &                  0 &        1 &            0 \\
Marietta &   2144 &                2 &                58417.608911 &                          20.14 &                          20.28 &                  256 &                   52 & 10.96 & 11.42 &      8.33e-03 &      1.77e-02 & 0.21 &        NaN &                  0 &        1 &            1 \\
Marietta &   2144 &                3 &                58846.631123 &                          19.45 &                          19.45 &                  309 &                  126 & 11.08 & 11.41 &      8.92e-03 &      1.27e-02 & 0.21 &        NaN &                  0 &        1 &            1 \\
Marietta &   2144 &                4 &                59300.621863 &                          20.58 &                          20.47 &                  410 &                   90 & 11.02 & 11.41 &      1.28e-02 &      1.39e-02 & 0.21 &   1.58e-02 &                  1 &        1 &            0 \\
 Yatskiv &   2728 &                0 &                57400.538519 &                          22.19 &                          21.77 &                   49 &                   75 & 12.42 & 12.56 &      1.66e-02 &      2.09e-02 & 0.11 &        NaN &                  0 &        1 &            1 \\
 Yatskiv &   2728 &                1 &                57826.587072 &                          28.61 &                          27.77 &                  113 &                   13 & 12.29 & 12.59 &      1.07e-02 &      1.65e-02 & 0.11 &        NaN &                  0 &        1 &            0 \\
 Yatskiv &   2728 &                2 &                58385.616541 &                          20.64 &                          19.43 &                  131 &                   28 & 12.30 & 12.65 &      1.23e-02 &      3.43e-02 & 0.11 &        NaN &                  0 &        1 &            0 \\
 Yatskiv &   2728 &                3 &                58790.634282 &                          23.23 &                          23.01 &                  231 &                   66 & 12.50 & 12.63 &      1.24e-02 &      1.82e-02 & 0.11 &        NaN &                  0 &        1 &            1 \\
 Yatskiv &   2728 &                4 &                59276.616956 &                          29.29 &                          29.21 &                  404 &                   85 & 12.31 & 12.56 &      1.55e-02 &      1.09e-02 & 0.11 &   1.35e-02 &                  1 &        1 &            1 \\
Osipovia &   4986 &                0 &                57591.583406 &                          27.46 &                          27.37 &                   48 &                   46 & 13.21 & 13.63 &      1.53e-02 &      1.65e-02 & 0.30 &        NaN &                  0 &        1 &            1 \\
Osipovia &   4986 &                1 &                58037.632049 &                          24.71 &                          23.86 &                  158 &                   37 & 13.23 & 13.63 &      1.01e-02 &      2.07e-02 & 0.30 &        NaN &                  0 &        1 &            1 \\
Osipovia &   4986 &                2 &                58515.672604 &                            NaN &                            NaN &                  NaN &                  NaN &   NaN &   NaN &           NaN &           NaN &  NaN &        NaN &                NaN &        0 &            0 \\
Osipovia &   4986 &                3 &                58974.596053 &                          25.95 &                          25.79 &                  191 &                   96 & 13.19 & 13.58 &      1.82e-02 &      9.53e-03 & 0.30 &   2.31e-02 &                  1 &        1 &            1 \\
Osipovia &   4986 &                4 &                59493.612303 &                          26.30 &                          24.83 &                  110 &                   35 & 13.22 & 13.60 &      1.10e-02 &      2.15e-02 & 0.30 &        NaN &                  0 &        1 &            0 \\
 1991 EN &   9857 &                0 &                57461.642500 &                            NaN &                            NaN &                  NaN &                  NaN &   NaN &   NaN &           NaN &           NaN &  NaN &        NaN &                NaN &        0 &            0 \\
 1991 EN &   9857 &                1 &                57882.583924 &                           9.72 &                           8.05 &                   36 &                    7 & 10.04 & 10.48 &      2.03e-02 &      6.54e-02 & 0.16 &        NaN &                  0 &        0 &            0 \\
 1991 EN &   9857 &                2 &                58269.545450 &                          11.38 &                          11.37 &                  140 &                   45 & 10.16 & 10.49 &      1.03e-02 &      1.41e-02 & 0.16 &        NaN &                  0 &        1 &            1 \\
 1991 EN &   9857 &                3 &                58613.616782 &                          11.57 &                          11.45 &                  255 &                   66 & 10.09 & 10.40 &      4.74e-03 &      1.36e-02 & 0.16 &        NaN &                  0 &        1 &            1 \\
 1991 EN &   9857 &                4 &                58997.589051 &                          11.40 &                          11.39 &                  209 &                  109 & 10.09 & 10.44 &      3.96e-03 &      7.08e-03 & 0.16 &        NaN &                  0 &        1 &            1 \\
 1991 EN &   9857 &                5 &                59414.593958 &                          11.17 &                          10.99 &                  352 &                   46 & 10.10 & 10.46 &      1.36e-02 &      1.34e-02 & 0.16 &   2.14e-02 &                  1 &        1 &            1 \\
\bottomrule
\end{tabular}

\caption{Results of the phase curve fitting algorithm to determine phase curves for each apparition of the selected objects from table \ref{tab:data_overview}.
NaN entries indicate where a phase curve fit could not be determined (e.g.\ the apparition was within the galactic plane and all observations were rejected).
We assign an index to each apparition for an object ($n_\text{app}$), where the first observation in an apparition was made at time $t_{\text{app},0}$.
The maximum phase angle observed in an apparition is given by $\text{max}(\alpha_{\text{app},o})$ and $\text{max}(\alpha_{\text{app},c})$ for the $o$ and $c$-filters respectively.
The number of detections used to fit the phase curve (after outlier rejection) is denoted by $N_{\text{fit},o}$ and $N_{\text{fit},c}$.
We show the fitted phase curve parameters for the $HG$ \protect\citetalias{bowellApplicationPhotometricModels1989} system, alongside their uncertainties ($\sigma_H$, $\sigma_G$) in both $o$ and $c$-filters.
The ``Primary Apparition'' flag indicates which apparition was fitted for both absolute magnitude and phase parameter.
We also flag whether an apparition passed the quality control checks (sections \ref{sec:phase_curve_fits} \& \ref{sec:phase_curve_classification}) in the $o$ filter (``Keep$_o$'') or if it passed in both $o$ and $c$ (``Keep$_{c-o}$'').
Only selected columns are displayed here.
The table of apparitions 
	is available in its entirety online as supplementary data and \href{https://doi.org/10.17034/db29dc08-5e21-403f-93f3-d60073b97916}{at this link},
with additional columns describing in detail the phase curve fit to each apparition for both the \citetalias{bowellApplicationPhotometricModels1989} and \citetalias{penttilaG1G2Photometric2016} models.
}
\label{tab:all_apparitions} 
\end{table*}

\begin{table*}  
\begin{tabular}{lrrrrrrrrrrr}
\toprule
    Name & Number & $N_{\textrm{app},o}$ & $F_\textrm{app}$ & $H_o$ & $H_c$ & $\sigma_{Ho}$ & $\sigma_{Hc}$ &  $G$ & $\sigma_G$ & $N_{\textrm{app,}c,o}$ & $H_c-H_o$ \\
\midrule
Moguntia &    766 &                    4 &             5.90 &  9.68 & 10.08 &      2.42e-03 &      1.22e-03 & 0.21 &   2.87e-03 &                      3 &      0.38 \\
 Priamus &    884 &                    3 &             0.79 &  8.43 &  8.86 &      9.36e-03 &      9.19e-03 & 0.23 &   1.48e-02 &                      3 &      0.43 \\
Marietta &   2144 &                    5 &             0.30 & 11.02 & 11.41 &      1.28e-02 &      1.39e-02 & 0.21 &   1.58e-02 &                      3 &      0.43 \\
 Yatskiv &   2728 &                    5 &             0.64 & 12.31 & 12.59 &      1.55e-02 &      1.09e-02 & 0.11 &   1.35e-02 &                      3 &      0.15 \\
Osipovia &   4986 &                    4 &             0.14 & 13.21 & 13.62 &      1.82e-02 &      9.53e-03 & 0.30 &   2.31e-02 &                      3 &      0.42 \\
 1991 EN &   9857 &                    4 &             0.37 & 10.10 & 10.45 &      1.36e-02 &      1.34e-02 & 0.16 &   2.14e-02 &                      4 &      0.35 \\
\bottomrule
\end{tabular}

\caption{The final \protect\citetalias{bowellApplicationPhotometricModels1989} phase curve parameters for the selected objects in tables \ref{tab:data_overview}, as determined from the individual apparitions in table \ref{tab:all_apparitions}.
Column $N_{\text{app},o}$ provides the number of apparitions in the $o$ filter that were used and $F_\text{app}$ is the parameter measuring the strength of the apparition effect (equation \ref{eqn:app_check}).
The overall absolute magnitude, phase parameter and the associated uncertainties for each object are presented as calculated using the methods in section \ref{sec:app_abs-mag_colours}.
We also provide the number of apparitions that had high quality phase curves in both filters ($N_{\text{app},c,o}$), for which we could calculate an asteroid colour $H_c-H_o$.
Available online.
}
\label{tab:B89_phase_curves} 
\end{table*}

\begin{table*}  
\begin{tabular}{lrrrrrrrrrrr}
\toprule
    Name & Number & $N_{\textrm{app},o}$ & $F_\textrm{app}$ & $H_{12o}$ & $H_{12c}$ & $\sigma_{H12o}$ & $\sigma_{H12c}$ & $G_{12}$ & $\sigma_{G12}$ & $N_{\textrm{app,}c,o}$ & $H_{12c}-H_{12o}$ \\
\midrule
Moguntia &    766 &                    4 &             5.87 &      9.63 &     10.02 &        1.19e-03 &        1.28e-03 &     0.43 &       8.52e-03 &                      2 &              0.37 \\
 Priamus &    884 &                    3 &             0.74 &      8.40 &      8.83 &        1.01e-02 &        9.45e-03 &     0.69 &       6.22e-02 &                      3 &              0.43 \\
Marietta &   2144 &                    4 &             0.35 &     10.96 &     11.37 &        7.70e-03 &        1.33e-02 &     0.63 &       6.01e-02 &                      4 &              0.41 \\
 Yatskiv &   2728 &                    5 &             0.62 &     12.36 &     12.63 &        6.24e-03 &        1.07e-02 &     0.73 &       4.17e-02 &                      2 &              0.18 \\
Osipovia &   4986 &                    4 &             0.14 &     13.07 &     13.47 &        8.80e-03 &        9.46e-03 &     0.18 &       5.59e-02 &                      3 &              0.42 \\
 1991 EN &   9857 &                    4 &             0.37 &     10.12 &     10.47 &        1.49e-02 &        1.35e-02 &     0.74 &       8.47e-02 &                      4 &              0.35 \\
\bottomrule
\end{tabular}

\caption{The final $H_{12}G_{12}$ \citetalias{penttilaG1G2Photometric2016} phase curve parameters for the selected objects from table \ref{tab:data_overview}, as determined from the individual apparitions in table \ref{tab:all_apparitions}.
These columns have the same meanings as in table \ref{tab:B89_phase_curves}, but now for \protect\citetalias{penttilaG1G2Photometric2016} phase curves as opposed to \protect\citetalias{bowellApplicationPhotometricModels1989}.
Available online.
}
\label{tab:P16_phase_curves} 
\end{table*}

The result of the analysis above are two databases. We have a database of phase curve fits by apparition derived for 427,662 MBA and Jupiter Trojan asteroids with \ATLAS observations gathered by \rockAtlas.
A sample of these results is presented in table \ref{tab:all_apparitions}.
We also have the Selected \ATLAS Phase-curves (SAP) database 
	containing the subset of filtered high quality fits for 106,055 unique asteroids; 93,373 and 91,103 in the \citetalias{bowellApplicationPhotometricModels1989} and \citetalias{penttilaG1G2Photometric2016} systems respectively).
	Approximately 9\% of these objects display strong apparition effects; 8450 \citepalias{bowellApplicationPhotometricModels1989} and 8131 \citepalias{penttilaG1G2Photometric2016}. 
	Samples of these datasets are provided in tables \ref{tab:B89_phase_curves} and \ref{tab:P16_phase_curves}.
In addition, we obtained $H_c-H_o$ colours for objects with suitable absolute magnitudes in both filters, resulting in colours for 66143 asteroids in the \citetalias{bowellApplicationPhotometricModels1989} system (64808 for \citetalias{penttilaG1G2Photometric2016}).
Please see the Data Availability statement at the end of this manuscript.


\begin{figure*}
\noindent
\begin{tabular}{@{}lll}
    \includegraphics{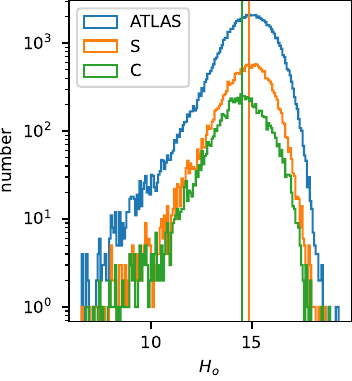} \hspace{-4pt}
    \includegraphics{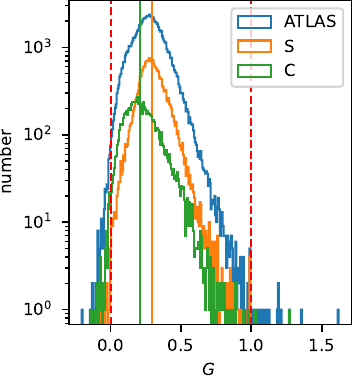} \hspace{-4pt}
    \includegraphics{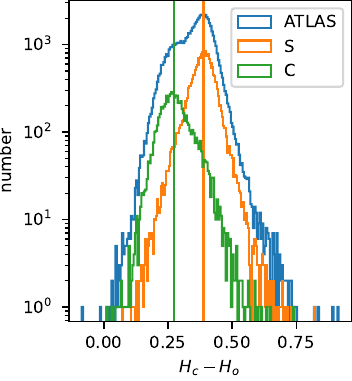}
\end{tabular}
    \caption{
    Log-scale distributions of $H$ (left), $G$ (middle) and $H_c-H_o$ colour (right) for \ATLAS phase curve fits that have passed the quality control checks (section \ref{sec:phase_curve_classification}) and are included in the SAP dataset (section \ref{sec:phase_curve_database}).
    $H$ and $G$ are given in the $o$-filter and \protect\citetalias{bowellApplicationPhotometricModels1989} model.
    In both figures we show the underlying distributions of S and C type asteroids within the \ATLAS dataset \protect\citep[taxonomy from \ssodnet;][]{berthierSsODNetSolarSystem2023}.
    The median values of the S and C-type distributions are indicated by the vertical line of matching colour and their values are as follows: $H_o(\mathrm{S}) = 14.83$, $H_o(\mathrm{C}) = 14.52$, $G(\mathrm{S}) = 0.29$, $G(\mathrm{C}) = 0.21$, $H_c-H_o(\mathrm{S}) = 0.39$, $H_c-H_o(\mathrm{C}) = 0.27$.
    In the middle panel we indicate the boundaries on $G$ for a monotonically increasing \protect\citetalias{bowellApplicationPhotometricModels1989} phase curve with vertical dashed lines.
    }
    \label{fig:phase_curve_HG_c-o_B89}
\end{figure*}

In figure \ref{fig:phase_curve_HG_c-o_B89} we present the distributions of $H_o$, $G$ and $H_c-H_o$ colour for asteroids observed by \ATLAS in the SAP database.
We also show 
the distributions of asteroids that have either an S-type or C-type taxonomic classification from the Solar system Open Database Network \citep[\ssodnet;][]{berthierSsODNetSolarSystem2023}, as these are the most common asteroid compositional classes.
While the overall $H$ distribution is smooth with a single dominant peak, the underlying S and C distributions peak at different values. 
This is an observational bias; the S-types have higher albedos than the darker C-types and are found more commonly in the inner asteroid belt \citep{usuiALBEDOPROPERTIESMAIN2013,demeoSolarSystemEvolution2014}.
Therefore the S-types are generally brighter than C types and they are better represented in \ATLAS observations down to fainter absolute magnitudes.
\\

The selection procedure made in section \ref{sec:phase_curve_classification} has successfully constrained the majority of the $G$ distribution to ``physical'' solutions between 0 and 1.
Only a small number of phase curves lie outside these limits ($0.3\%$ of the dataset) however we did not wish to arbitrarily exclude them or other phase curves near the boundaries from our results.
Visual inspection of these particular objects show that the fitted phase curves are consistent with the observations, albeit generally for a  more limited range of phase angles.
For $G$ outside the $(0,1)$ range it is usually not until higher phase angles than those observed by \ATLAS that the phase function becomes non-monotonic/asymptotic.

As $G$ decreases the slope of the approximately linear component at phase angles $\alpha \gtrsim 5^\circ $ gets steeper.
In particular the outliers with $G \lesssim 1$ have phase curves that are entirely compatible with the observations.
On the other hand, as $G$ increases the slope at larger phase angles decreases and the model fails to describe the opposition effect at phase angles $<5\, \si{deg}$.
Our handful of objects with $G \gtrsim 1$ have relatively flat phase curves over the observed phase angle range and with rotational variation this allows for an unphysical phase curve to be fitted, where brightness decreases at opposition.
However, the vast majority of objects in our dataset have sensible phase curve fits and we can see a mildly bimodal distribution in $G$ caused by the underlying distributions of S and C-type phase parameters.
This bimodality primarily arises due to the different albedos (and compositions) of the two taxonomic classes; the brighter S-type asteroids have a mean geometric albedo of 0.21 whereas the dark C-types have typical albedos of 0.07 \citep{usuiALBEDOPROPERTIESMAIN2013}.
As shown by \cite{belskayaOppositionEffectAsteroids2000a} the gradient of the phase curve at large phase angles is correlated with albedo.
In the \citetalias{bowellApplicationPhotometricModels1989} model, the higher median value of $G$ for the S-type asteroids leads to a flatter phase function, which corresponds to a higher albedo as expected.

\begin{figure}
    \includegraphics{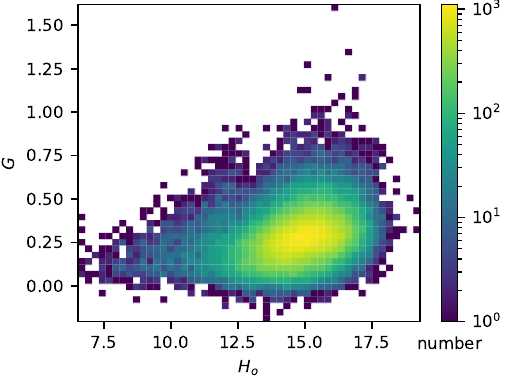}    
    \caption{
    Two dimensional histogram which shows the variation of $G$ with absolute magnitude $H$ for asteroids in the SAP dataset.
    There is a positive correlation between $H$ and $G$, with an increasing dispersion in $G$ for fainter objects (increasing $H$).
    }
    \label{fig:HG_B89}
\end{figure}

There is a correlation between $H$ and the range of $G$ we find in the data (figure \ref{fig:HG_B89}).
This is partly due to the enhanced presence of S-type asteroids as $H$ increases, as we are most sensitive to detecting small asteroids in the inner main belt where S-types are most populous, and these objects typically have higher values of $G$.
However the main cause is a systematic effect; as $H$ increases asteroid size is decreasing and we are generally considering fainter photometric observations with higher uncertainty.
This increased uncertainty means there is more scatter in fitted phase parameters for fainter objects with larger $H$.
This $H$-$G$ correlation driven by photometric uncertainty acts in addition to any uncertainty caused by rotational lightcurve effects.
For example, analysis of the $HG$ dataset published by \citetalias{schemelZwickyTransientFacility2021} shows a similar trend.

\begin{figure}
    \includegraphics{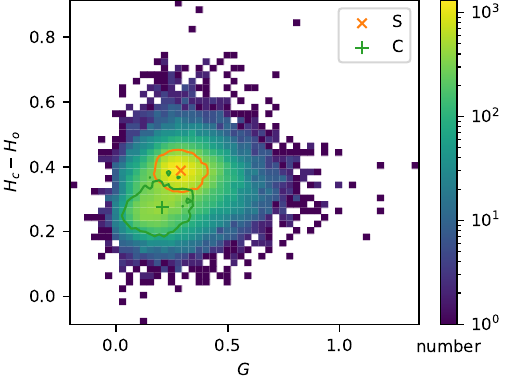}
    \caption{
    Two dimensional histogram of $H_c-H_o$ colour vs.\ $G$.
The bimodal distribution of matched S and C-type asteroids \protect\citep[\ssodnet;][]{berthierSsODNetSolarSystem2023} is highlighted by the contoured regions, which indicates the 25\% level with respect to the largest histogram bin value of their sub-distributions in $H_c-H_o$ vs.\ $G$ parameter space.
The median values of the S and C-types with \ATLAS colours are $G=0.29,0.21$ and $H_c-H_o = 0.39,0.27\, \si{mag}$ respectively.
    }
    \label{fig:H_c-o_G_B89}
\end{figure}

Figure \ref{fig:phase_curve_HG_c-o_B89} shows that there is a strong bimodality for the $H_c-H_o$ colour distribution in the SAP database. 
As described in section \ref{sec:app_abs-mag_colours} the data presented is for the subset of objects that have observations of sufficient quality in both \ATLAS filters for one or more apparitions (with $F_\text{app}<0.9$).
The bimodality is again driven by the different surface compositions of the dominant S and C type asteroids.
We obtain median $H_c-H_o$ colours of $0.39\pm0.05$, $0.27\pm0.06$ (uncertainties of one standard deviation) for the objects in our dataset matched with S and C-type classifications respectively \citep{berthierSsODNetSolarSystem2023}.
These values are compatible with the expected $H_c-H_o$ colours calculated by \cite{erasmusInvestigatingTaxonomicDiversity2020},
who convolved the mean Bus-DeMeo spectra \citep{demeoExtensionBusAsteroid2009} with the \ATLAS filter response to obtain $H_c-H_o$ colours of 0.388 and 0.249 for S and C-type asteroids respectively.
Figure \ref{fig:H_c-o_G_B89} shows the 2-D bimodal distribution of $H_c-H_o$ vs.\ $G$ in the SAP database. 
Taken together, the phase parameter and colour can be extremely valuable in distinguishing between the dominant S and C-type asteroid compositions (illustrated further in section \ref{sec:asteroid_families}).

\begin{figure}
    \includegraphics{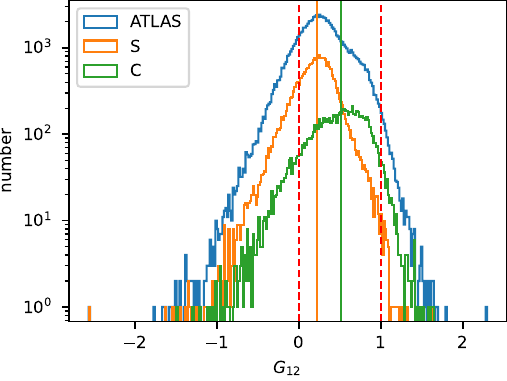}
    \caption{
    Log-scale distribution of \protect\citetalias{penttilaG1G2Photometric2016} $G_{12}$ phase parameters for the quality controlled SAP dataset.
Distributions of objects with S and C-type taxonomic classifications from \protect\cite{berthierSsODNetSolarSystem2023} are also shown.
    Similar to figure \ref{fig:phase_curve_HG_c-o_B89} the median values of the S and C-type distributions are indicated by coloured vertical lines (with values of $G_{12}(\mathrm{S}) = 0.22$,
    	$G_{12}(\mathrm{C}) = 0.51$) 
    and the limits $0<G_{12}<1$ for a ``physical'' phase curve are shown as vertical dashed lines.
    }
    \label{fig:phase_curve_G12_P16}
\end{figure}

\begin{figure}
\centering
\noindent
\begin{tabular}{@{} c}
	\includegraphics{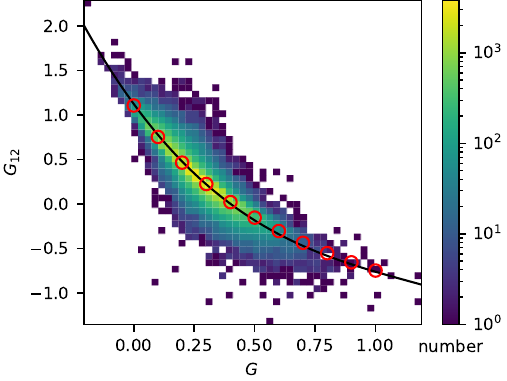} \\ 
	\includegraphics{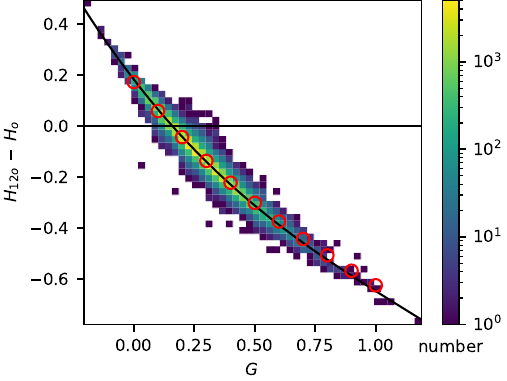}
\end{tabular}
    \caption{
Two dimensional histogram showing the relation between the $G_{12}$ and $G$ parameters for objects in our database (upper panel).
A 3rd-order polynomial is fitted to the data as shown by the black curve (equation \ref{eqn:G12_G_polyrelation}).
The circular markers indicate the values of $G_{12}$ when a $H_{12}G_{12}$ model is fitted to a $HG$ model (appendix \ref{appendix:compare_HG_HG12}).
This empirical test follows the trend in the \ATLAS data closely.
The bottom panel shows the two dimensional histogram of the difference in fitted absolute magnitude $H_{12}-H$ vs.\ $G$.
Again, the curve is a 3rd-order polynomial fit (equation \ref{eqn:H12_G_polyrelation}) and the circular markers indicate the expected difference in absolute magnitude when $H_{12}G_{12}$ is fitted to $HG$.
The horizontal line indicates the zero value of the $H_{12}-H$ axis.
    }
    \label{fig:HG_HG12_relations}
\end{figure}

We have also derived the phase curve parameters of asteroids using the \citetalias{penttilaG1G2Photometric2016} $H_{12}G_{12}$ system, where we use the notation $H_{12}$ to denote absolute magnitudes in this system.
As described in section \ref{sec:phase_curve_classification} the quality control cuts were developed on the $HG$ dataset and then applied separately to the residuals of the $H_{12}G_{12}$ phase curve fits. Therefore not all asteroids are present in both datasets.
Regardless, both phase curve models show broadly similar results.
As before the phase parameter and colour distributions are both bimodal due to the dominant  S and C-type asteroids, the former shown in figure \ref{fig:phase_curve_G12_P16}.
The quality control metrics developed for $HG$ do not lead to as strong restrictions on $G_{12}$ for ``physical'' values between 0 and 1.
This appears to be mostly due to inherent differences between the $HG$ and $H_{12}G_{12}$ models.
Figure \ref{fig:HG_HG12_relations} shows that there is a strong correlation between $G_{12}$ and $G$ for objects in our dataset with a phase curve fit in both models.
We have verified that this is caused by the different formulations of the two models and is not due to systematic effects in our phase curve fitting procedure.
As detailed in appendix \ref{appendix:compare_HG_HG12} (figure \ref{fig:phase_curves_G_G12}), we generated a range of ideal model $HG$ phase curves with $0\leq G\ \leq1$ and then fitted those simulated observations with the $H_{12}G_{12}$ model.
The locus of best fit $G_{12}$ values for phase curves with parameter $G$ is shown in the top panel of figure \ref{fig:HG_HG12_relations}.
It can be seen that fitting \citetalias{penttilaG1G2Photometric2016} to observed phase curves naturally gives rise to a range of $G_{12}$ larger than $(0,1)$.

\begin{figure}
    \includegraphics{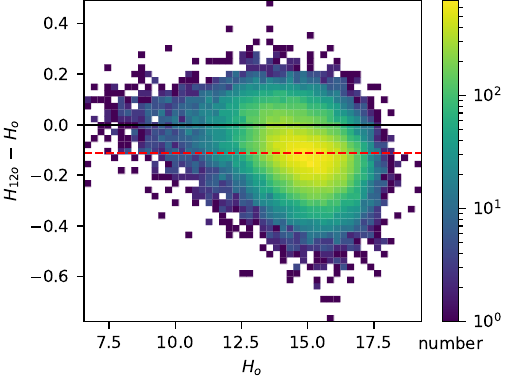}
    \caption{
    Two dimensional histogram which shows the variation of $\Delta H=H_{12} - H$ with $H$.
    The median difference in absolute magnitude between the two phase curve models is $-0.11$ (indicated by the horizontal dashed line). There is a negative correlation between $\Delta H$ and $H$, as well as a trend for a larger dispersion in $\Delta H$ for fainter objects (see also figure \ref{fig:HG_HG12_relations}).
    The horizontal solid line indicates the zero value of the $H_{12}-H$ axis.
    }
    \label{fig:H_P16-H_B89}
\end{figure}

Additionally we find an offset between the absolute magnitudes in the two systems as shown in figure \ref{fig:H_P16-H_B89}.
This is similar to the results of \citetalias{mahlkeAsteroidPhaseCurves2021} (their figure 4).
There is a trend of increasing offset in $H$ for objects with fainter absolute magnitudes.
This means that applying the \citetalias{penttilaG1G2Photometric2016} model results in lower values of $H$ (i.e.\ brighter, larger objects) than for the corresponding \citetalias{bowellApplicationPhotometricModels1989} fit.
We confirm that this effect also arises due to the inherent differences between the two models.
\citetalias{penttilaG1G2Photometric2016} is more sensitive at modelling opposition effects and has a third basis function which accounts for non-linear brightness surges at low phase angles (equation \ref{eqn:P16_model}), which can lead to brighter values of $H$.
In the test described above the exact $HG$ phase curve extends all the way to zero phase.
The $H_{12}G_{12}$ model cannot match $HG$ for both the approximately linear behaviour at large phase angles and the non-linear opposition brightening at low phase angles (figure \ref{fig:phase_curves_G_G12}).
The $HG$ and $H_{12}G_{12}$ models show the most significant divergence at low phase angles which explains the differences in absolute magnitude between the two models.

We show the relation between $H_{12}-H$ and phase parameter $G$ for the phase curves in our database in the lower panel of figure \ref{fig:HG_HG12_relations}. 
We note that it is also well described by a locus derived from fitting \citetalias{bowellApplicationPhotometricModels1989} phase curves with the \citetalias{penttilaG1G2Photometric2016} system (see again
figure \ref{fig:phase_curves_G_G12} in appendix \ref{appendix:compare_HG_HG12}). 
Therefore the shift in absolute magnitude is an inherent property of comparing two different phase curve models rather than being caused by the phase angle coverage or photometric uncertainty of \ATLAS data. 
Clearly the \citetalias{penttilaG1G2Photometric2016} model more accurately captures the opposition surge which is present for many asteroids, however, for historical reasons the \citetalias{bowellApplicationPhotometricModels1989} model is often still in use (e.g.\ MPC, JPL, \astorb).
This $H$ difference is systematic across both $o$ and $c$-filter phase curve fits, hence the resulting $H_c-H_o$ colours for both models are practically indistinguishable. 
In the \citetalias{penttilaG1G2Photometric2016} system we find median $H_c-H_o$ colours that are nearly identical to the \citetalias{bowellApplicationPhotometricModels1989} system: $0.39 \pm 0.05$ and $0.27 \pm 0.06$ for S and C-types respectively (uncertainties of one standard deviation).

\subsection{Comparison with previous studies} \label{sec:comparison_to_previous_work}

\begin{figure}
	\centering
	\noindent
    \begin{tabular}{@{} c}
         \includegraphics{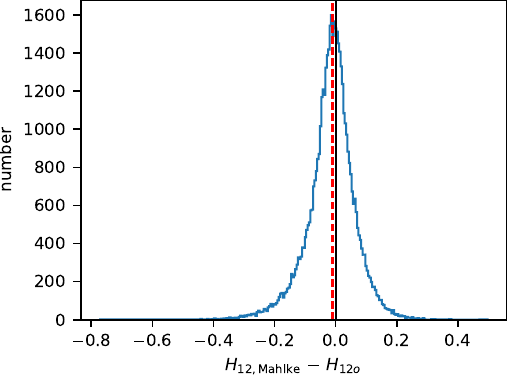}\\
         \includegraphics{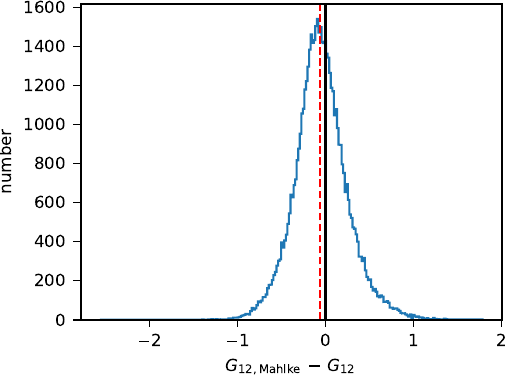}
    \end{tabular}
    \caption{
    Histograms showing the difference in $H$ (upper panel) and $G_{12}$ (lower panel) of our phase curve fits in the \protect\citetalias{penttilaG1G2Photometric2016} system, cross-matched with the results of \protect\cite{mahlkeAsteroidPhaseCurves2021} who also used \ATLAS data.
    The median of each distribution is indicated by the vertical dashed line with values of -0.01 and -0.06 magnitudes for the upper and lower panels respectively.
    The vertical solid lines mark the zero value of the $x$-axis.
    }
    \label{fig:compare_mahlke_HG12}
\end{figure}

The most direct comparison of our results to published work is with the study by \citetalias{mahlkeAsteroidPhaseCurves2021}, who also made use of \ATLAS data for asteroid phase curve analysis.
They utilised the photometry for $\sim180,000$ asteroids in the period of 2015-2019.
From this they determined phase curves with the \cite{muinonenThreeparameterMagnitudePhase2010a} $HG_1G_2$ and \citetalias{penttilaG1G2Photometric2016} $H_{12}G_{12}$ models for $\sim94000$ asteroids using MCMC Bayesian parameter inference, with fits in both $o$ and $c$ filters.
Similar to our methods they did not directly account for rotational effects in their phase curve model.
However they did estimate the influence of rotational and apparition driven magnitude variations on the uncertainties of their phase curve fits.
In this work we considered the phase curves of separate apparitions and performed more rigorous rejection of outlying photometry.
Another important difference in methodology is that we did not restrict phase curve parameters when fitting, whereas \citetalias{mahlkeAsteroidPhaseCurves2021} restricted MCMC walkers to bounds of $0 < G_{12} < 1$ to ensure monotonically increasing phase curves.
This leads to a pile up of $G_{12}$ values at the boundaries as seen in figure 5 of \citetalias{mahlkeAsteroidPhaseCurves2021}.
Despite these differences there is reasonable agreement for absolute magnitudes of objects that are present in both datasets (for 49,558 matched objects); most objects also have similar values of $G_{12}$ (figure \ref{fig:compare_mahlke_HG12}).
The median difference in absolute magnitude $H_{12,\text{Mahlke}} - H_{12} = -0.01 \pm 0.08\, \si{mag}$ (one standard deviation) and the difference in $G_{12}$ is also relatively near zero with a median $G_{12,\text{Mahlke}} - G_{12} = -0.06 $ albeit with a higher standard deviation of $0.31$.

In another study \cite{mcneillComparisonPhysicalProperties2021}, hereafter \citetalias{mcneillComparisonPhysicalProperties2021}, published phase curve properties and colours of 342 Jupiter Trojans asteroids using \ATLAS observations from 2015-2018.
They fitted $HG$ \citetalias{bowellApplicationPhotometricModels1989} phase curves to the $o$ and $c$-filter data, accounting for rotational brightness variation by repeatedly fitting to a randomised 50\% subsample of observations.
In this work we considered using a similar approach, however we found that it did not significantly change the estimates of $HG$ (or their uncertainties).
This is because a sub-sample of a phase curve exhibiting strong rotational effects will have similar residuals to the original data.
The results of \citetalias{mcneillComparisonPhysicalProperties2021} provide the $G$ parameter and $c-o$ colour of their sample, which we were able to match with 317 objects in our science dataset (253 with colours).
Despite using more data from the same survey, not all objects in the \citetalias{mcneillComparisonPhysicalProperties2021} sample pass our quality control checks.
We applied a robust methodology to reject outlying photometry and assessed the quality of phase curve fits to individual apparitions.
Furthermore we rejected any objects displaying strong apparition effects from our colour analysis.

\begin{figure}
\includegraphics{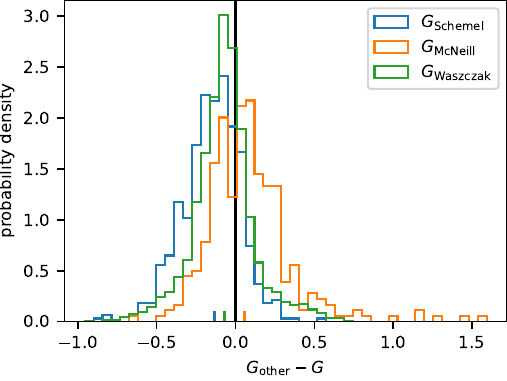}
\caption{
    Histogram distributions showing the difference in $G$ \protect\citepalias{bowellApplicationPhotometricModels1989} for phase curve fits in our SAP database cross-matched with the previous results of \protect\cite{waszczakASTEROIDLIGHTCURVES2015}, \protect\cite{mcneillComparisonPhysicalProperties2021} and \protect\cite{schemelZwickyTransientFacility2021}.
    Each distribution is shown as a probability density to better compare datasets with significantly different numbers of objects.
    The coloured $x$-axis ticks indicate the median value of the corresponding distribution, whereas the vertical black line indicates the zero value of the $x$-axis.
    The median offset for each distribution is $G_\mathrm{Schemel}-G=-0.13$, $G_\mathrm{McNeill}-G=0.06$, $G_\mathrm{Waszczak}-G=-0.07$.
    }
    \label{fig:compare_schemel_mcneill_G}
\end{figure}

\begin{figure}
    \includegraphics{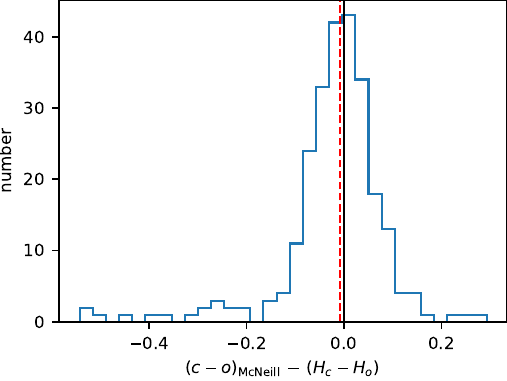}
    \caption{
    Histogram comparing the \ATLAS $H_c-H_o$ colour distribution of objects in this work cross-matched with the previous colours ($c-o$) published by \protect\cite{mcneillComparisonPhysicalProperties2021}, who also investigated the properties of the Jupiter Trojans using \ATLAS data.
    The median value of the distribution is -0.008 magnitudes (vertical dashed line), the vertical solid line indicates the zero value of the $x$-axis.
    }
    \label{fig:compare_mcneill_c-o}
\end{figure}

When we compare the matched objects we find a median phase parameter difference of  $G_{\text{McNeill}} - G = 0.06$, with quite a large standard deviation of 0.28 (figure \ref{fig:compare_schemel_mcneill_G}).
Similarly for the difference in $H_c-H_o$ colours, the median difference is $-0.008 \pm 0.1$ (figure \ref{fig:compare_mcneill_c-o}).
These difference distributions are centred relatively close to zero, however, the spread in values between the two studies should not be surprising.
These differences primarily arise due to the differing methodology described above, furthermore, we have made use of additional \ATLAS observations that were taken since their analysis was performed.
In particular, the dataset of \citetalias{mcneillComparisonPhysicalProperties2021} contains a small number of objects with $1< G <2$ for which we obtain more realistic values of $G$.

Considering another study focused on the Jupiter Trojans, \citetalias{schemelZwickyTransientFacility2021} determined $HG$ \citetalias{bowellApplicationPhotometricModels1989} phase curves for a significant number of Jupiter Trojans asteroids ($1049$ objects) using ZTF data \citep{bellmZwickyTransientFacility2019}.
Importantly this study accounts for rotational lightcurve variation by including the probability of an observation being made at a particular phase for a given amplitude and building a likelihood model for the apparent magnitude of each observation.
They determine posterior distributions for $H$, $G$ and $g-r$ colour, which more accurately capture the true uncertainty in these parameters rather than the formal fitted uncertainties quoted in this work.
Furthermore they account for offsets in $H$ and changing rotational amplitude for different apparitions of an asteroid.
We match Jupiter Trojans that appear in both our study and that of \citetalias{schemelZwickyTransientFacility2021}, for a total of 571 cross-matched objects.

There is reasonable agreement between $H$ given that we have not accounted for rotational effects, the median difference in $H_\text{Schemel}-H$ is $0.07 \pm 0.13$. 
However, there is a significant offset in $G$ distributions: $-0.13 \pm 0.18$ (figure \ref{fig:compare_schemel_mcneill_G}).
This means that in general our work finds larger values of $G$ for Jupiter Trojans, which corresponds to flatter phase curves (at phase angles $>5\, \si{deg}$ in the approximately linear regime).
This arises due to the increased photometric uncertainty of \ATLAS observations for the dark and distant Jupiter Trojans, the objects with the greatest semimajor axes in our study.
This uncertainty, combined with rotational brightness variation means that the \ATLAS phase curves have a large scatter around the expected phase curve behaviour, giving a flatter appearance.
Our simple methodology tends to find solutions with higher $G$ when such scatter is significant.
This seems to be a common effect when rotational effects are not accounted for.
A recent study by \cite{donaldsonCharacterizingNucleusComet2023} showed that the phase curve of the Jupiter family comet nucleus 162P/Siding-Spring (which has low albedo, similar to the Jupiter Trojans) got steeper when the rotational brightness variation was subtracted from the phase curve.
Furthermore, it would appear that the scatter introduced by rotational variation (and also photometric uncertainty) tends to ``disguise'' the opposition brightness surge of an asteroid at low phase angles.
		However if rotational effects were accounted for one would expect the opposition surge (if present) to be more accurately resolved.
In this case one should fit the phase curve with a model such as \citetalias{penttilaG1G2Photometric2016} that better resolves this non-linear surge when compared to \citetalias{bowellApplicationPhotometricModels1989} (figure \ref{fig:phase_curves_G_G12}).

We also consider the work of \cite{waszczakASTEROIDLIGHTCURVES2015} who simultaneously determined rotation periods and phase curves of asteroids observed by the Palomar Transient Factory \citep{lawPalomarTransientFactory2009} in $R$ and $g$-filters, resulting in a dataset of $\sim9000$ objects.
They determined phase curves in the two parameter \cite{shevchenkoAnalysisAsteroidBrightnessPhase1997} model, $HG$ \citetalias{bowellApplicationPhotometricModels1989} and the two parameter model of \cite{muinonenThreeparameterMagnitudePhase2010a} (the precursor to \citetalias{penttilaG1G2Photometric2016}).
Therefore we compare their \citetalias{bowellApplicationPhotometricModels1989} phase curve parameters to ours for 5199 MBAs and Jupiter Trojans common to both datasets (figure \ref{fig:compare_schemel_mcneill_G}).
The median difference in $G$ and its standard deviation, $-0.07 \pm 0.20$,  are of similar magnitude to the difference with \citetalias{mcneillComparisonPhysicalProperties2021}.
However both \cite{waszczakASTEROIDLIGHTCURVES2015} and \citetalias{schemelZwickyTransientFacility2021} $\Delta G$ distributions are shifted towards negative values.
These works accounted for rotational variation in their phase curve models, and for most objects obtain lower values of $G$ compared to this work and that of \citetalias{mcneillComparisonPhysicalProperties2021}.
This again implies that magnitude dispersion due to rotation and photometric uncertainty tends to flatten the fitted phase curve and push the distribution of $G$ to higher values.

\section{DISCUSSION} \label{sec:discussion}

\subsection{The Selected \ATLAS Phase-curves database (SAP)}

The SAP database contains absolute magnitudes and phase curve parameters in the \citetalias{bowellApplicationPhotometricModels1989} and/or \citetalias{penttilaG1G2Photometric2016} systems for 106055 main belt and Trojan asteroids. The median measured value of $H_o$ changes from $16.6\, \si{mag}$ at the inner edge of the main belt at $a=1.9$~au to $13.0\, \si{mag}$ for the Hilda asteroids at $a=4$~au, and to $11.6\, \si{mag}$ for the Jupiter Trojan asteroids at $a=5.2$~au. 
The median uncertainties of the non-linear least squares fits for all phase curves in the SAP database are $\sigma_G = 0.04$ and $\sigma_{G12} = 0.11$. 
In section \ref{sec:asteroid_families} we examine the phase parameter and colour properties of asteroid families in the SAP database.
Then in section \ref{sec:results-jupiter_trojans} we describe an initial analysis of Jupiter Trojan asteroids using these data.

\subsection{Asteroid families in the SAP database}
\label{sec:asteroid_families}

\begin{figure}
\centering
\noindent
\begin{tabular}{@{} c}
    \includegraphics{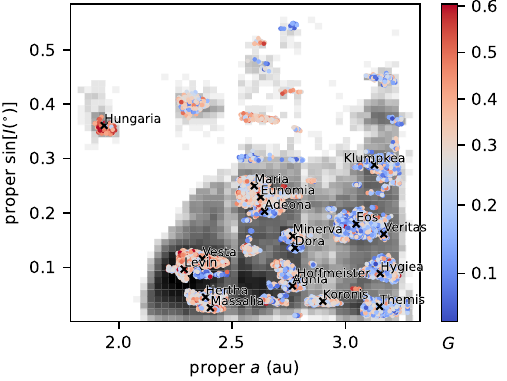}\\
    \includegraphics{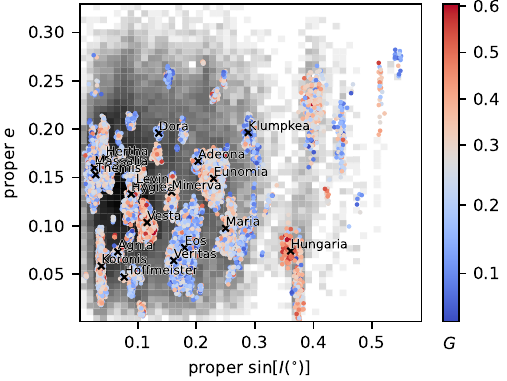}
\end{tabular}
\caption{
    Proper orbital element distributions of MBAs in the SAP database (elements from \astdys).
    The proper semimajor axis is plotted against the sine of proper inclination (upper panel) and proper eccentricity (lower panel).
    The 63,197 non-family members in the dataset are displayed as a two dimensional histogram in greyscale. 
    The 25,479 asteroids that are a member of a family (as classified by \astdys) are colour-coded by $G$ with a diverging colour map, where the colour transition is centred on the median value of $G$.
    The same colour scheme is used for both plots and objects with phase parameters outside the $1-99\%$ quantile of the family member $G$ distribution were excluded from the plot in order to avoid stretching of the colour map by the small number of more extreme values.
    As such we only display asteroid family members with $0.03<G<0.60$ (where the median $G$ of this sample is 0.26).
    Some of the larger asteroid families are labelled by crosses at their mean proper elements \protect\citep{milaniAsteroidFamiliesClassification2014}. 
    }
\label{fig:orbital_proper_elements_diagram}
\end{figure}

To illustrate the usefulness of the SAP database, in figure \ref{fig:orbital_proper_elements_diagram} we show the \citetalias{bowellApplicationPhotometricModels1989} phase parameters as a function of proper orbital elements (obtained from the \astdys database\footnote{\url{https://newton.spacedys.com/astdys/} accessed 3rd January 2023}), highlighting the asteroid family members using the \astdys classifications. 
In this figure we have indicated the presence of non-family MBAs using a greyscale two dimensional histogram and the dynamical family members are overplotted as coloured markers, appearing as clusters in proper element space.
The marker colour in figure \ref{fig:orbital_proper_elements_diagram} provides the phase parameter $G$ of each asteroid and many of the dynamical families clearly share similar phase parameters.
This indicates a shared albedo/composition amongst family members, which is to be expected when a family is created from the catastrophic collisional fragmentation of a parent asteroid. 
There is a general decrease in the $G$ of asteroid families as proper semimajor axis increases, which is driven primarily by the increasing fraction of C-type asteroids in the outer main belt.

\begin{figure*}
	\centering
		\includegraphics{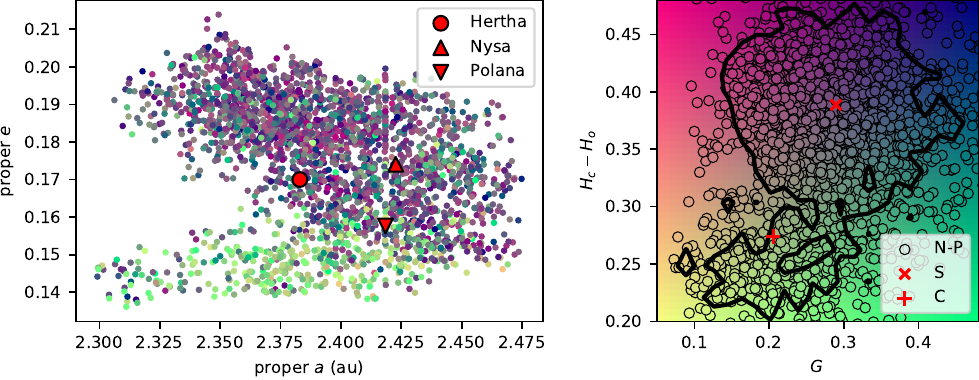}
	\caption{Here we show the \astdys proper orbital element distribution of Nysa-Polana (N-P) asteroid family members in the SAP dataset, in proper eccentricity vs.\ proper semimajor axis space (left panel).
		We indicate the proper elements of the main asteroids of the complex (Nysa, Polana and Hertha) with shaped markers.
		Every other family member is colour coded according to a two dimensional colour map, $H_c-H_o$ colour against phase parameter $G$.
		This colour map is displayed in the right-hand panel, where we have also marked the locations of the SAP Nysa-Polana family members in this parameter space (circular markers).
		The contour indicates the 10\% level relative to the maximum count of the underlying two dimensional histogram of the distribution.
		We have also indicated the median $H_c-H_o$ and $G$ of the S and C-type asteroids in the full SAP database (cross and plus markers respectively, values from figure \ref{fig:phase_curve_HG_c-o_B89}).
	}
	\label{fig:Nysa-Polana-H_c-H_o_G_B89}
\end{figure*}

We examine in more detail the Nysa-Polana asteroid family, which has previously been identified as having a complex dynamical structure.
It is considered to be a large ``clan'' grouping also containing the low-numbered asteroid Hertha, for which the group may also be named \citep{cellinoPuzzlingCaseNysa2001,milaniAsteroidFamiliesClassification2014}.
In addition to the sub-structure of the group in proper element space it has been shown that this large family also contains a mixture of asteroid taxonomies.
\cite{erasmusInvestigatingTaxonomicDiversity2020}, also using \ATLAS data, derived probabilities of an S-like vs.\ C-like composition for Nysa-Polana members using their own measurements of the $H_c-H_o$ colour. 
They showed that there is a clear two-part taxonomic mixture within this dynamical grouping.
Here we build upon this study by considering the combination of the $H_c-H_o$ colour and phase parameter $G$ derived in this work for identifying taxonomy.
Figure \ref{fig:Nysa-Polana-H_c-H_o_G_B89} (left) shows the proper orbital elements of the Nysa-Polana (Hertha) family members that are present in the SAP database.
The presence of two clusters in proper element space is clear and to indicate the taxonomic diversity of these objects we have used a two dimensional mapping of $H_c-H_o$ vs.\ $G$ to colour each marker.
This colour map is displayed in the right-hand panel of figure \ref{fig:Nysa-Polana-H_c-H_o_G_B89} and it illustrates that this combination of $H_c-H_o$ and $G$ can provide a powerful tool for identifying broad taxonomic information of asteroids observed by sparse, wide field surveys.
We have indicated the distribution of Nysa-Polana members in this $H_c-H_o$ vs.\ $G$ space and indicated the location of the median S and C-type parameters as measured from the full SAP database (figure \ref{fig:phase_curve_HG_c-o_B89}).
This analysis shows that there is strong correlation between dynamical clustering and taxonomic clustering in the Nysa-Polana group.
These clusters have distinct S-like and C-like features and supports the theory that this group is made up of nested families with distinct parent bodies.
Further analysis by the community of the dynamical and collisional families present in the SAP dataset is encouraged.

\subsection{Jupiter Trojan asteroids in the SAP database} \label{sec:results-jupiter_trojans}

Another important dynamical population to investigate with the SAP dataset are the low-albedo, optically red Trojan asteroids in 1:1 mean-motion resonance with Jupiter at orbital semimajor axes around $5.2 \, \si{AU}$ \citep{marzariOriginEvolutionTrojan2002}. 
The Jupiter Trojans are distributed in two groups that are leading/trailing Jupiter by approximately $60\, \si{degrees}$, defined by the L4/L5 Lagrange points of the restricted three body problem.
In the Nice model of planetary migration \citep{gomesOriginCataclysmicLate2005,tsiganisOriginOrbitalArchitecture2005,morbidelliChaoticCaptureJupiter2005} the Trojans formed exterior to the giant planets and were scattered and captured onto their current orbits as the giant planets dynamically evolved.
Alongside $N$-body simulations supporting this theory, observations of Jupiter Trojan surface composition has shown these objects to be more similar to outer Solar System objects than the inner small body population \citep{wongHYPOTHESISCOLORBIMODALITY2016}.
There are over $10,000$ known Jupiter Trojans \citep[as listed by the MPC, complete down to around $H=14\, \si{mag}$,][]{hendlerObservationalCompletionLimit2020}, and determination of the underlying size distribution indicates a number asymmetry of $\sim 1.4$ between the L4/L5 groups \citep{gravWISENEOWISEOBSERVATIONS2011}.
Within the Nice model framework this could be caused by depletion of the trailing group by the ejection of a massive planet from the early inner Solar System \citep{nesvornyCAPTURETROJANSJUMPING2013}.
Therefore there is interest in seeing how similar (or not) the Jupiter Trojan groups are in terms of their physical properties as well.
This will help to determine if they indeed share the same parent population and whether the number asymmetry has affected their properties through collisional processing since depletion.
Remote observations of Jupiter Trojans are particularly relevant given the recent launch of the LUCY mission \citep{levisonLucyMissionTrojan2021}.
This mission will obtain in-situ observations of Jupiter Trojans in both dynamical groups, which will provide a valuable anchor-point to the studies made using disk-unresolved photometry.

The Jupiter Trojans have been well observed by the \ATLAS survey, providing a substantial and consistent photometric dataset with which to study these objects.
Notably, the Trojans have the lowest maximum phase angle ($\text{max}(\alpha) \leq 11^\circ$) in our database as they are the most distant objects we have included in this study (figure \ref{fig:phase_angle_max}). 
This naturally limits the phase angle coverage, although  we can still sample any opposition surge for many of them, with observations of most Jupiter Trojans typically reaching a phase angle of $1\, \si{deg}$.  
We note that in 2017 the leading L4 cloud was at low galactic latitude and has since moved out of the galactic plane, whereas the trailing L5 cloud has since been approaching lower galactic latitudes. 
Hence both Jupiter Trojan clouds have been sampled in \ATLAS observations whilst away from the galactic plane, albeit over different time periods, thanks to the long baseline of the survey.

\begin{figure*}
\centering
    \includegraphics{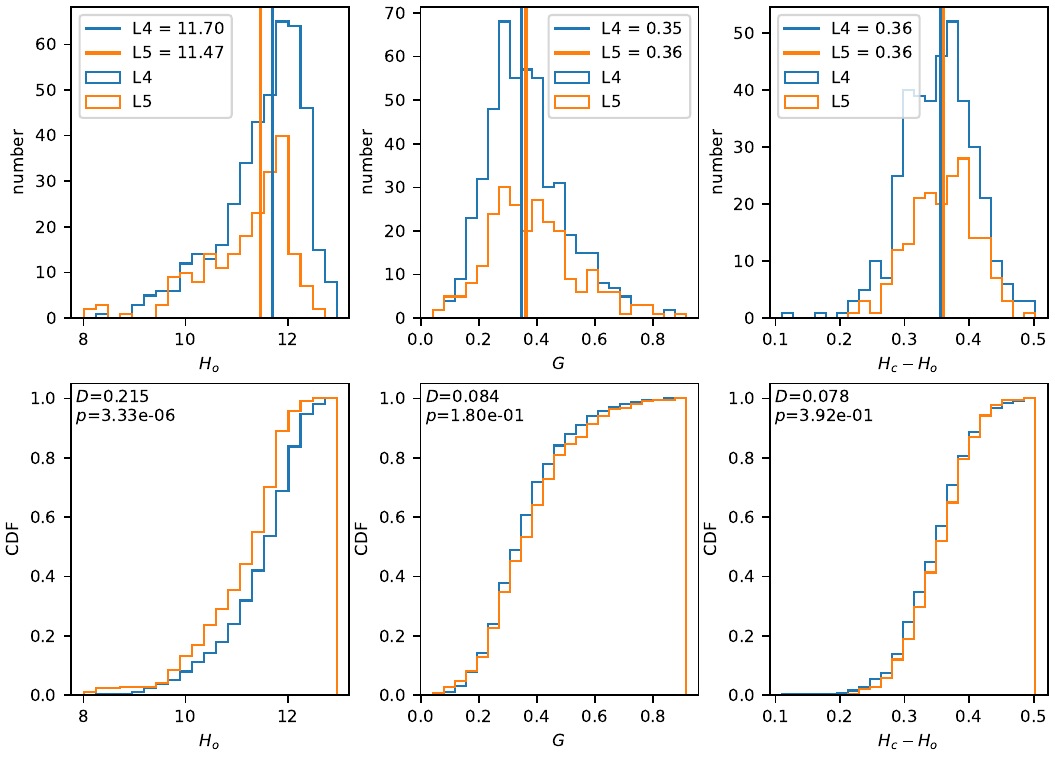}
    \caption{
    Histogram distributions of $H$, $G$ and $H_c-H_o$ colour, in the \protect\citetalias{bowellApplicationPhotometricModels1989} system, for the L4 and L5 Jupiter Trojans in the SAP dataset (top row, left, middle and right panels respectively).
    The distributions for the L4 and L5 groups are shown as coloured histograms; the median of each distribution is indicated by the matching colour vertical line.
    The cumulative distribution functions (CDFs) of each parameter are shown in the bottom row, allowing us to compare the distributions via the two-sample Kolmogorov-Smirnov test.
    The KS statistic $D$ is determined from the maximum difference between the CDFs at a given value and when this is sufficiently large (at a level determined by the $p$-value) we can reject the null hypothesis that the two datasets are drawn from the same underlying distribution.
    }
    \label{fig:jupiter_trojan_L4_L5_KS}
\end{figure*}

Here we consider the phase curve properties of the Jupiter Trojans present in our SAP dataset. 
This contains 740 Jupiter Trojans with high quality $HG$ fits in the $o$-filter, of which 570 also have good fits in the $c$-filter suitable for calculating a $H_c-H_o$ colour (section \ref{sec:app_abs-mag_colours}). 
103/740 of these Jupiter Trojans display strong apparition effects and our sample contains 1.98 more Jupiter Trojans at L4 than L5.
This is similar to the L4/L5 number ratio for observed Jupiter Trojans \citep[for counts of Jupiter Trojans recorded by the MPC,][]{liAsymmetryNumberL42023} but is larger than the number ratio of $\sim 1.4$ when the underlying size-frequency distributions are accounted for \citep{gravWISENEOWISEOBSERVATIONS2011,uehataSizeDistributionSmall2022a}. 
We compare our distributions in absolute magnitude, $H$, of these two groups (figure \ref{fig:jupiter_trojan_L4_L5_KS}, left panels) using a two sample Kolmogorov-Smirnov (KS) test.
In a two sample KS test the parameter $D$ describes the maximum distance (at a given value) between the cumulative distributions of the two datasets being tested. 
The KS statistic is calculated with a $p$-value which indicates the probability that the null hypothesis is true, i.e.\ that both datasets were drawn from the same underlying distribution. 
Conventionally a result is statistically significant when $p<0.05$ (see also equation \ref{eqn:KS_D_crit}).
For the L4/L5 $H$ distributions this test results in a KS statistic of $D=0.2$ (with a $p$-value of $3\e{-6}$) from which we reject the null hypothesis that the L4 and L5 absolute magnitudes are drawn from the same distribution.
There is no strong evidence in the literature that the L4, L5 groups should have different size distributions \citep[e.g.][]{uehataSizeDistributionSmall2022a}.
In this case we determined that the differences in $H$ were introduced into our study through the initial watchlist of asteroids monitored by \rockAtlas.
There is no significant difference in the absolute magnitude distribution for the L4/L5 Trojans when using $H_V$ from the full \astorb dataset.
However when we compare the \astorb $H_V$ of only the L4/L5 Trojans which were selected for the \rockAtlas watchlist, a KS test displays statistically significant differences similar to those seen in figure \ref{fig:jupiter_trojan_L4_L5_KS} (left).
Therefore we can rule out observational bias between the L4 and L5 groups in the \ATLAS survey as the main cause of this effect, and attribute it instead to selection bias.

We performed similar tests on the remaining parameters retrieved from the SAP dataset.
For the \citetalias{bowellApplicationPhotometricModels1989} phase parameter $G$ the two sample KS test results are $D = 0.08$ and $p = 0.18$.
Likewise for $H_c-H_o$ colours (for the smaller number of objects with good fits in both $o$ and $c$), the KS test results are $D = 0.08$ and $p = 0.39$ (figure \ref{fig:jupiter_trojan_L4_L5_KS}, middle and right panels respectively).
These $D$ values are near the boundary at which we would consider rejecting the null hypothesis.
To ensure that these results are not influenced by small number statistics or dominated by specific objects in the dataset we performed repeated KS tests on bootstrapped samples of each parameter.
We assessed the variation in $D$ for the resampled data, the results of which are detailed further in appendix \ref{appendix:Jupiter_Trojan_KS_tests} (figure \ref{fig:jupiter_trojan_KS_bootstrap}).
Overall these tests confirm that there is no statistical evidence for any differences in the distribution of phase parameter $G$ or $H_c-H_o$ colour between the L4 and L5 Jupiter Trojans.
This is in contrast to the reported dissimilarity in $c-o$ colours of the L4/L5 Jupiter Trojans by \citetalias{mcneillComparisonPhysicalProperties2021}.
This difference in results is most likely due to the shorter \ATLAS survey baseline at the time of their study, which meant that \citetalias{mcneillComparisonPhysicalProperties2021} had fewer observations available for phase curve analysis.
Furthermore, although the statistically different $H$ distributions are due to the \rockAtlas watchlist, in this work we have not thoroughly accounted for observational biases between L4 and L5 in the \ATLAS dataset.
Biases could have been introduced by observations of the L4/L5 groups as they are moving away from/towards the galactic plane over the course of the survey.
We can see in figure \ref{fig:jupiter_trojan_L4_L5_KS} that the \ATLAS observations for the L4 group generally go down to fainter absolute magnitudes.
Therefore there could be size effects on $G$ and $H_c-H_o$ in addition to the greater phase curve fitting uncertainty for fainter objects (figure \ref{fig:HG_B89}).

As a sanity check we repeated the above analysis but instead using the Jupiter Trojan dataset published by \citetalias{schemelZwickyTransientFacility2021}.
At first we found that the KS tests for the L4/L5 $H$ and $G$ distributions seemed to indicate statistically significant differences (high $D$ values with low $p$).
However these results are highly dependent on the quality of the phase curve fit.
\citetalias{schemelZwickyTransientFacility2021} report uncertainties derived from the posterior distributions of their parameters, which are generally larger than those reported in works like our own, but which are a better representation of the true range of parameters compatible with the observations.
When we consider only their phase curves with better than median uncertainties \citepalias[using the same methods as][in their work]{schemelZwickyTransientFacility2021} then any differences between the L4 and L5 groups become statistically insignificant, similar to our results (figure \ref{fig:schemel_jupiter_trojan_KS_bootstrap}).

\begin{figure}
	\centering
	\noindent
    \begin{tabular}{@{} c}
    \includegraphics{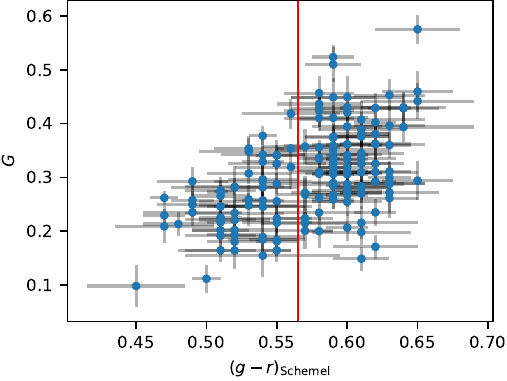}\\
         \includegraphics{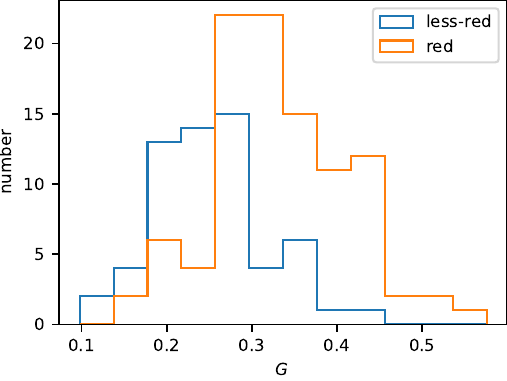}
    \end{tabular}
    \caption{
Scatter plot of phase parameter $G$ from the SAP dataset against the ZTF $g-r$ colours from \protect\cite{schemelZwickyTransientFacility2021}, for objects matched between the two datasets (upper panel).
    Similar to their analysis we only include objects from both datasets with lower than median uncertainties (compare to their figure 6).
    The vertical line corresponds to the colour boundary at $g-r=0.565$ which divides the ``red'' and ``less-red'' Jupiter Trojan compositional classes
    The lower panel shows the histogram distributions of \ATLAS $G$ for the red and less-red Jupiter Trojans from the panel above.
    }
    \label{fig:schemel_g-r_GB89}
\end{figure}

Although the two dynamical clouds have similar distributions of phase curve parameter, it has been shown by several authors that the Jupiter Trojans in both clouds can be split into two compositional groups based on their colours and/or spectra; the less-red and red sub-classes \citep{emeryNEARINFRAREDSPECTROSCOPYTROJAN2011,szaboPropertiesJovianTrojan2007}. 
 Recently, it has been found by \citetalias{schemelZwickyTransientFacility2021} that these sub-classes show distinct phase curves at optical wavelengths.
 As indicated in figure \ref{fig:schemel_g-r_GB89}, this bimodality in phase parameter $G$ is also present in our dataset.  
In this figure we have plotted the phase parameter $G$ obtained from this work against the $g-r$ colour derived from ZTF data by \citetalias{schemelZwickyTransientFacility2021}, using only objects with better than median uncertainties in both datasets \citepalias[replicating their analysis, see figure 6 of][]{schemelZwickyTransientFacility2021}. Of the 571 asteroids common to both datasets, 159 meet this requirement.
Similar to the previous work we find a strong positive correlation between our $G$ values and their $g-r$ colours, with a Spearman rank correlation coefficient of $\rho = 0.51$ ($p=7\e{-12}$).
We then divided objects into red and less-red groups by a ZTF colour boundary of $g-r=0.565$ and analysed the corresponding distributions of $G$.
We replicate the results of \citetalias{schemelZwickyTransientFacility2021} whereby less-red Jupiter Trojans generally have lower $G$ than red Jupiter Trojans, albeit our phase parameters are uniformly shifted to higher values as discussed previously in section \ref{sec:comparison_to_previous_work}.
A KS test of the two distributions prove that they are indeed distinct ($D=0.48$, $p=2\e{-8}$).

We also search for correlation between $G$ and $H_c-H_o$ colour for the 570 Jupiter Trojans with colours in the \ATLAS dataset.
The Spearman rank correlation coefficient of these parameters is $\rho=0.07$ ($p = 0.08$), which indicates that there is no significant correlation. 
This is due to the large width of the \ATLAS $o$ and $c$-filters which overlap around the spectral ``kink'' which distinguishes the red and less-red types \citep{emeryNEARINFRAREDSPECTROSCOPYTROJAN2011, mcneillComparisonPhysicalProperties2021}. 

In summary, our database of \ATLAS phase curves shows no significant differences in phase parameter $G$ or $H_c-H_o$ colour between the L4 and L5 Jupiter Trojans.
This is in general agreement with previous studies and supports the consensus that both groups of Jupiter Trojans were captured from a common source population.
Although there is evidence for a difference in the average shapes of the L4/L5 groups \citepalias[presumably due to differences in number densities and collisional environment,][]{mcneillComparisonPhysicalProperties2021}, \ATLAS data shows no corresponding difference in their surface properties caused by different collision histories.
Furthermore, although \ATLAS $H_c-H_o$ colour can be a useful diagnostic when distinguishing between the S and C-type asteroids \citep[figure \ref{fig:H_c-o_G_B89} and also][]{erasmusInvestigatingTaxonomicDiversity2020} these broad filters are less effective at differentiating more subtle spectral features, such as the red and less-red Jupiter Trojans.
Regardless, the phase parameter $G$ captures information on the differences in surface properties of the two Jupiter Trojan classes.
When combined with more detailed colour information from other surveys (figure \ref{fig:schemel_g-r_GB89}) a stronger distinction between the red and less-red Jupiter Trojans can be inferred, highlighting the benefits of utilising information from multiple sources.

\subsection{Limitations of the database}

\subsubsection{Outlying photometry}
In order to reliably and automatically fit phase curves to a large number of objects we developed various selection criteria to reject outlying observations before and during the phase fitting algorithm (section \ref{sec:phase_curve_fits}).
It is necessary to remove any false matches caused by the automated nature of the \rockAtlas pipeline, however, we risk throwing away valid measurements, e.g.\ increases in asteroid brightness due to activity/outburst or rotation/apparition effects.
Previous work such as \citetalias{mahlkeAsteroidPhaseCurves2021} addressed this by conducting a straight $1\, \si{mag}$ difference cut between the predicted and observed magnitudes.
We found that a combination of data cuts based on magnitude difference and standard deviation of residuals provided a more robust method of outlier rejection, which we confirmed through a process of visual inspection of the extreme outliers in the resulting dataset.

Furthermore, we considered additional properties of each observation.
We rejected all data points within $10^\circ$ of the galactic plane as these measurements have a high probability of stellar contamination.
In some cases this led to entire apparitions being rejected as an object moved through the galactic plane.
On one hand such a rough cut could throw away valid measurements, on the other when the Milky Way thickness exceeds our $20^\circ$ span (e.g.\ near the galactic bulge) we may not be rejecting contaminated data.
However, we found that most galactic plane sources were rejected anyway by our strict $\theta<1 \arcsec$ constraint on \rockAtlas ephemeris matching.

Our methods for matching sources could be improved further by crossmatching all detections with stationary source catalogues in order to flag data points contaminated by nearby sources and to conduct a more precise rejection.
This is currently beyond the scope of this work, 
however, we have shown that our results are generally consistent with existing literature which also used \ATLAS data (section \ref{sec:comparison_to_previous_work}).
Therefore the \rockAtlas observation matching is not a significant source of uncertainty.

\subsubsection{Catalogue contamination by poor quality fits}

One of the complex aspects of this work was the production of a final high quality phase curve dataset.
We fitted all available apparitions and then ideally one should be able to use the uncertainties of the fitted phase curve parameters to assess the quality of the dataset.
However as discussed previously, we do not account for rotational brightness variation and therefore the formal uncertainties produced by the Levenberg-Marquardt algorithm do not reflect the true range of possible values for the phase curve parameters.
We therefore had to develop a series of cuts based on a variety of metrics which we tried to choose in a relatively agnostic way so as to not overly influence the final results (section \ref{sec:phase_curve_classification} and appendix \ref{appendix:quality_control_metrics}).
We did this by considering especially the phase curves of objects with unusually high values of $G$ and $\sigma_{G}$ and adjusting our metrics such that these truly unphysical fits were rejected.
We acknowledge that this approach relies on somewhat subjective decisions.
A more realistic estimate of the phase parameter uncertainties could have been obtained by methods in which rotational variation is accounted for \citepalias[e.g.][]{schemelZwickyTransientFacility2021}.
However, this would add further computational cost to calculating phase parameters for each apparition of a large number of objects.
We did consider ``quick'' solutions to deal with rotational effects, e.g.\ binning the observed magnitude over some timescale or by fitting random samples of the observations, however we found that these methods did not significantly improve the quality of the final dataset.

\subsubsection{Phase curve wavelength dependence}

We determined the phase parameter ($G$, $G_{12}$) for each object from its primary apparition in the $o$-filter, and kept this phase parameter fixed when fitting all other apparitions.
There is evidence that phase curve properties vary with wavelength in a ``phase-coloring'' effect \citep{alvarez-candalAbsoluteColorsPhase2022,durechAsteroidModelsReconstructed2020}. 
However in the \ATLAS dataset the coverage in the $c$-filter is much sparser compared to the $o$-filter. 
In most cases we lacked sufficient data to obtain an independent phase curve for the $c$-filter data.
In order to obtain usable colour information we were forced to assume the slope parameters to be wavelength independent in order to find $H_o$ and $H_c$. 
This is a reasonable assumption given the accuracy of \ATLAS photometry and not accounting for rotational variation. 
These effects probably dominate over any changes in phase parameter due to the wavelength of observation.

\subsubsection{Rotational lightcurve effects}

While developing our methodology we attempted several simple methods to incorporate rotational brightness variations into our phase curve fits.
For example we considered a Monte Carlo approach whereby each observation was re-sampled based on its measurement uncertainty and phase curves were repeatedly fit \citepalias[similar to][]{mcneillComparisonPhysicalProperties2021}.
We also attempted to average out rotational variation by binning reduced magnitudes nightly when fitting the phase curves.
However, we found that these methods did not adequately account for rotational variations in the lightcurves.
When resampling observations for objects displaying clear lightcurve variations the photometric uncertainty is generally small compared to the amplitude and so the fitted phase curve parameters and their uncertainties did not differ significantly from our original method.
Similarly, binning the data did not have much effect on the final parameters.
Furthermore, binning observations of objects with a wide possible range of rotational periods (minutes to days) compared to the \ATLAS observational cadence would likely introduce a number of additional biases to our analysis.
A more robust treatment would account for the unknown rotational phase and amplitude of each observation and so the true uncertainties of the phase curve parameters would be much wider but reflect better the true distribution of possible parameters (e.g.\ \citet{waszczakASTEROIDLIGHTCURVES2015,alvarez-candalPhaseCurvesSmall2022}, \citetalias{schemelZwickyTransientFacility2021}).
However such methods are generally more computationally intensive to implement, therefore in this work we stuck with a simple phase curve fitting method, neglecting rotational effects as described in section \ref{sec:phase_curve_fits}.
This allowed us to consider a sizeable sample of asteroids, and their individual apparitions, in a timely manner.
As such, the formal fit uncertainties produced in this work should be used to compare relative goodness of fit within the dataset; they do not necessarily reflect the true uncertainty in the fitted phase curve parameters if rotational effects are strong.

\subsection{Applicability to future surveys}

In this work we have demonstrated that sparse wide field survey data can be exploited to investigate and broadly compare the properties of different asteroid populations.
In particular, by considering the phase parameters and surface colours derived in this work one may compare and contrast different dynamical groups such as the asteroid families and Jupiter Trojans.
However as discussed above, in order to make more precise assessments one must take into account the rotational variation introduced by the asteroid lightcurve, either in a statistical sense (e.g.\ \citet{waszczakASTEROIDLIGHTCURVES2015,alvarez-candalPhaseCurvesSmall2022}, \citetalias{schemelZwickyTransientFacility2021}) or with more detailed lightcurve shape modelling \citep[e.g.][]{durechAsteroidModelsReconstructed2020}.
Despite our simple phase curve model, we have built upon previous phase curve studies by considering the phase curve properties of individual apparitions for a large sample of over $100,000$ asteroids in the Main Asteroid Belt to Jupiter Trojan region, with data spanning a significant temporal baseline since 2017.
By doing so we have been able to identify objects that are undergoing changes in observed absolute magnitude between apparitions.
By excluding these objects from our science analysis of absolute magnitudes and colours we have improved the purity of our sample.

This variation in brightness known as the apparition effect is driven by the shape and spin state of an object and means that closer study is required in order to determine its true absolute magnitude.
Studies such as \cite{durechAsteroidModelsReconstructed2020} have shown that \ATLAS sparse photometry can be analysed using lightcurve inversion techniques, although in most cases additional dense lightcurve observations are required to reduce the parameter space to be searched.
Of the 9168 objects we have identified with strong apparition effects (in either \citetalias{bowellApplicationPhotometricModels1989} or \citetalias{penttilaG1G2Photometric2016}) only 2828 have reported periods in \ssodnet \citep[][accessed 15th March 2024]{berthierSsODNetSolarSystem2023}.
Therefore this list of objects with demonstrated lightcurve effects makes for promising targets for future dedicated observations in order to accurately define their physical properties.

Assessing apparition effects informs us of the elongation and spin pole orientation of an object.
If we could disentangle the spin pole orientation for a large database of objects showing apparition effects we could assess the competing effects of collisional vs.\ YORP evolution on the spin pole \citep{pravecAsteroidRotations2002}.
This could be achieved by searching for relations between asteroid size and spin pole orientation, whereby the spin pole of asteroids is expected to evolve away from the orbital plane at a rate which is inversely proportional to size \citep[e.g.][]{cibulkovaDistributionSpinaxesLongitudes2016}.
Apparition effect objects could also be relevant for comparing shapes and/or collisional evolution, in particular for the L4/L5 Jupiter Trojans which show indications of different shape distributions \citepalias{mcneillComparisonPhysicalProperties2021}.

The presence of the aforementioned apparition effects must be accounted for when analysing the results of wide field surveys expected to observe large numbers of objects over a long baseline.
In particular, LSST will discover approximately $6\e{6}$ MBAs over 10 years, with 100s of observations for most objects.
In addition to moving object discovery, LSST will provide automatically derived properties of observed asteroids such as lightcurves, phase curves and colours \citep{lsstsciencecollaborationLSSTScienceBook2009}. 
Given the volume and depth of the survey it will be difficult to followup the expected large numbers of faint asteroids, therefore LSST will likely be the only source of data for many of these objects.
Considering the apparition effect is therefore essential in order to determine accurate absolute magnitudes, which directly effects studies regarding the size distribution of small Solar System bodies, such as the formation and evolution of planetesimals, properties of asteroid families, planetary defence, etc.

Furthermore, our approach which has involved detailed analysis of phase curve residuals offers a framework to monitor for unusual observations that might indicate an object of interest.
We considered a range of statistics for the phase curve residuals, and in the vast majority of cases any rejected apparitions were due to poor observational coverage or low quality photometry.
However, certain physical processes may also lead to deviations from the expected phase curve behaviour, e.g.\ rapid changes in viewing geometry \citep{jacksonEffectAspectChanges2022a} or sudden increases in brightness due to the onset of activity \citep{jewittAsteroidCometContinuum2022a}.
For example, asteroid 2008 GO98 was found to be active in July 2017 \citep{kokhirovaResultsObservationsDualstatus2021}. 
Our methodology produced a most unusual phase curve for this object, with an extreme value of $G=1675$ which was determined from an apparition spanning from 8th June until 9th December 2017.
Future work in this area could involve identifying these outliers in phase curve parameter space and using the \ATLAS forced photometry server to get cutouts to search for activity. 
Continuous updates of derived physical properties (such as absolute magnitude, phase curve and colours) as new observations are made will be an essential part of large surveys such as LSST.
Therefore learning how to identify truly unusual objects in this process will greatly advance the study of relatively rare objects such as the active asteroids.

\section{Conclusions} \label{sec:conclusions}

We have compiled a database of \ATLAS asteroid photometry by matching the ephemerides of 4.5\e{5} asteroids with accurate orbital elements to the detections produced by the \ATLAS \dophot pipeline, using \rockAtlas \citep{Young_rockAtlas}.
We fitted phase curves in the $HG$, $H_{12}G_{12}$ models \citep{bowellApplicationPhotometricModels1989,penttilaG1G2Photometric2016} to the apparition in the $o$-filter with the best combination of number of observations and phase angle coverage.
We then fitted all other apparitions for $H$ with these derived phase parameters fixed.
This allowed us to assess any changes in $H$ during the survey and also to determine the $H_c-H_o$ surface colour of many asteroids.
We used various metrics to select 106,055 asteroids with well defined phase curve fits for the final Selected \ATLAS Phase-curves (SAP) dataset, which is available online.

Within the SAP dataset we identified around 9000 asteroids that show strong apparition effects, i.e.\ large shifts in $H$ over the long baseline of \ATLAS observations.
These objects most likely have elongated shapes and tilted spin poles in order to display these effects. 
We suggest such asteroids should be prioritised for followup lightcurve observations and analysis.

Our measured absolute magnitudes and phase curve parameters were compared to previous studies, and we found that our analysis which does not account for rotation is in general agreement with similar studies. However, our method fails to accurately reproduce the Jupiter Trojan $G$ distribution compared to more sophisticated methods that include rotational variation \citep{waszczakASTEROIDLIGHTCURVES2015,schemelZwickyTransientFacility2021}.
We found that by not taking into account rotational brightness variations when fitting the phase curves, the increased dispersion and larger photometric uncertainties for the faint Jupiter Trojans led to flatter phase curves in this work.

In spite of this, we found that the asteroid colours and phase parameters provided in the SAP database can be a useful tool for comparing the properties of various asteroid populations.
We demonstrated this for main belt asteroid families, whereby families can be distinguished by their differing distributions of $G$.
Our dataset illustrates the known trend towards lower values of $G$ as semimajor axis increases, caused by the increasing fraction of C-type asteroids in the outer main belt.
Additionally, we examined in closer detail the Nysa-Polana (Hertha) dynamical complex.
The SAP database allowed us to use both $H_c-H_o$ colour and $G$ to infer compositional differences of the family members.
We demonstrated that this combination of physical parameters traces the differing compositions within the dynamical sub-structure of the larger family group.
Our results support theories that the Nysa-Polana group is composed of nested families formed of at least two parent bodies, which had different S-like and C-like compositions.

We also conducted a study of the Jupiter Trojans in the SAP database, the dynamical and physical properties of which help to constrain the evolution of the Solar System and the migration of the planets in particular.
We found that there were no statistically significant differences between the L4 and L5 groups for the distributions of $G$ and $H_c-H_o$.
We did measure a discrepancy in the distribution of $H$ between the L4 and L5 groups, however, we determined that this resulted from an initial bias in asteroid selection. 
Furthermore, we found that the broad $o$ and $c$-filters could not uniquely distinguish between red and less-red Jupiter Trojan compositional classes, however, when combined with colour measurements from other sources \citep[e.g.\ ZTF $g-r$ colours;][]{schemelZwickyTransientFacility2021} the difference in \ATLAS phase parameter $G$ between the two classes was emphasised.
In general our results support the consensus that the Jupiter Trojans were captured from a common source population.

In this work we chose a phase curve model without rotational lightcurve variations to better investigate individual apparitions of a large number of asteroids. 
However, we recommend that future studies account for rotational effects, e.g.\ via likelihood models including rotational variation \citep[for sparse data e.g.][]{schemelZwickyTransientFacility2021} and/or full CLI techniques \citep[for objects with additional lightcurve rotation periods e.g.][]{durechAsteroidModelsReconstructed2020}.
Accurate phase curve modelling with realistic uncertainties on absolute magnitude and phase curve parameters are essential when comparing the optical properties of different small body populations.
This will be particularly relevant for large, sparse surveys such as LSST which will discover and characterise approximately 6\e{6} asteroids over its 10 year baseline 
		\citep{lsstsciencecollaborationLSSTScienceBook2009}.
Many LSST objects will be too faint and/or numerous for dedicated followup therefore extracting accurate parameters from sparse data alone, and quickly highlighting interesting outliers, is a pressing issue.

Author's note: Since submission we have been made aware of development of the \texttt{sHG$_{1}$G$_{2}$} model \citep{carryCombinedSpinOrientation2024}; i.e.\ a ``spinned'' version of the $HG_1G_2$ \cite{muinonenThreeparameterMagnitudePhase2010a} model. 
	This enhanced phase curve model also fits the oblateness and spin axis of an object in order to account for shape driven rotational effects.
Such a model will certainly help to address the long-standing issues in determining accurate phase curve parameters from sparse data.

\section*{Acknowledgements}

The authors wish to thank the reviewer, Josef {\v D}urech, for useful comments that helped to improve this manuscript.
This work has made use of data from the Asteroid Terrestrial-impact Last Alert System (ATLAS) project. ATLAS is primarily funded to search for near earth asteroids through NASA grants NN12AR55G, 80NSSC18K0284, and 80NSSC18K1575; byproducts of the NEO search include images and catalogs from the survey area.  The ATLAS science products have been made possible through the contributions of the University of Hawaii Institute for Astronomy, the Queen's University Belfast, the Space Telescope Science Institute, the South African Astronomical Observatory (SAAO), and the Millennium Institute of Astrophysics (MAS), Chile.
This research has made use of IMCCE's SsODNet VO service (\url{https://ssp.imcce.fr/webservices/ssodnet/}).
J.E.R. acknowledges support from STFC award ST/T00021X/1 and the Royal Society. 
A.F. acknowledges support from STFC award ST/T00021X/1. 
J.E.R. also acknowledges the support and guidance of Cyrielle Opitom, whose discussions helped improve the quality of the work.
For the purpose of open access, the author has applied a Creative Commons Attribution (CC BY) licence to any Author Accepted Manuscript version arising from this submission.
In addition to software cited in the paper, the following packages were also used in this work: 
\verb|matplotlib| \citep{hunterMatplotlib2DGraphics2007}, 
\verb|numpy| \citep{harrisArrayProgrammingNumPy2020}, 
\verb|scipy| \citep{virtanenSciPyFundamentalAlgorithms2020}, 
\verb|scikit-learn| \citep{pedregosaScikitlearnMachineLearning2011}, 
\verb|pandas| \citep{mckinneyDataStructuresStatistical2010}, 
\verb|pds4_tools| \citep{nagdimunov_pds4_tools}, 
\verb|pycolormap_2d| \citep{spinner_pycolormap}.

\section*{Data Availability}

The data bases constructed in this work are available in their entirety online as supplementary data or via the Queen’s University Belfast PURE open data repository (\url{https://doi.org/10.17034/db29dc08-5e21-403f-93f3-d60073b97916}). 
They may also be obtained via
reasonable e-mail request to the lead authors.



\bibliographystyle{mnras}
\bibliography{zotero_library,extra_refs}




\section*{Supporting Information}

Supplementary data are available at MNRAS online.
\\

\noindent \textbf{Table 1.} Overview of the observations made by ATLAS and collected
by \rockAtlas, for each asteroid considered in this study.

\noindent \textbf{Table 2.} Results of the phase curve fitting algorithm to determine
phase curves for each apparition of the objects from Table 1.

\noindent \textbf{Table 3.} The final \citetalias{bowellApplicationPhotometricModels1989} phase curve parameters for the selected objects
in Tables 1, as determined from the individual apparitions in Table 2.

\noindent \textbf{Table 4.} The final $H_{12}G_{12}$ \citetalias{penttilaG1G2Photometric2016} phase curve parameters for the selected
objects from Table 1, as determined from the individual apparitions
in Table 2.
\\

Please note: Oxford University Press is not responsible for the content
or functionality of any supporting materials supplied by the authors.
Any queries (other than missing material) should be directed to the
corresponding author for the article.

\appendix

\section{Metrics for high quality phase curves} \label{appendix:quality_control_metrics}

The temporal and phase angle coverage of an asteroid imaged by \ATLAS will vary depending on many factors, e.g.\ accessible viewing geometry relative to the Earth, sky location relative to bright objects (such as the Moon, Jupiter and Saturn) and/or missed observations due to weather/technical issues.
Likewise the quality of observations obtained by \dophot will depend on factors such as object brightness, photometric seeing and/or sky brightness.
Furthermore, the matching of \dophot detections to known asteroid ephemerides may lead to false matches with stellar sources in the high density galactic plane or glints near bright objects and other such artifacts.
Therefore, poor quality phase curves can occur when there is not enough observational coverage to adequately constrain the phase curve, photometric uncertainty is high and/or there is significant outlying photometric observations.

We performed data cuts on the observations of each asteroid to account for obvious photometric outliers (section \ref{sec:phase_curve_fits}), although our methods required sufficient data in each apparition such that the initial $HG(G=0.15)$ fit provided a reasonable description of the general phase curve behaviour.
Considering the individual apparitions of each asteroid reduces the number of observations used for each phase curve fit, as such after performing our phase curve fitting algorithm there will many apparitions which do not produce accurate phase curve fits.
Therefore we used a series of cuts based on various metrics describing the observational coverage and residuals of each phase curve fit to reject poor quality apparitions from our final dataset.
By analysing the residuals of the phase curve fit at each apparition we were able to identify and exclude individual apparitions for which an accurate phase curve could not be obtained.
Therefore one bad apparition did not necessarily lead to complete rejection of an object.
The primary apparition provides us with the fitted phase parameter ($G$ or $G_{12}$) for each asteroid, therefore these apparitions had somewhat stricter requirements on the number of data points and phase angle coverage.
The phase parameter was kept fixed when fitting the non-primary apparitions.
For these fits the most common phase curve error was unphysical shifts in $H$ caused by outlying photometry missed by our data selection, poor phase angle coverage, and/or large rotational variations driven by asteroid shape.
We found that these problematic apparitions were best characterised by the statistical properties of their residuals.
For a high quality phase curve fit of an asteroid with low rotational amplitude we would expect the residuals to be approximately normally distributed around a residual of 0 with low standard deviation\footnote{In reality the likelihood function of brightness for a sinusoidal lightcurve peaks at the maximum and minimum amplitude \citepalias[see figure 1 of][]{schemelZwickyTransientFacility2021} although this would not be clearly seen for low amplitude lightcurves given the typical accuracy of \ATLAS photometry.}.
As such the standard deviation and mean value of the residuals proved to be valuable metrics.
We also performed a Kolmogorov-Smirnov test to identify distributions of residuals that differed significantly from a normal distribution.
Furthermore, we used the Pearson correlation coefficient of the residuals as a function of phase angle as a metric. 
This metric should ideally be near zero (indicating no correlation) but will approach $1$ or $-1$ when there is strong correlation in the residuals.
Correlated residuals implies that the phase curve model was not consistent with the observations.

The values of each metric cut were chosen through a process of trial and error alongside detailed visual inspection of individual \citetalias{bowellApplicationPhotometricModels1989} phase curve fits.
As far as possible we tried to avoid choosing values in a way that directly influenced the final distributions of phase curve parameters, i.e.\ we did not simply cut phase curves with $G<0$ or $G>1$. 
We did however choose values that led to a distribution such that the vast majority ($97.7 \%$) of $HG$ fits had $0<G<1$.
This also removed the objects with extremely high uncertainties in $G$, while not removing an excessive number of objects from the final database.
We did have to perform a hard cut in $\sigma_{H}$, $\sigma_{G}$ as a final step, as there was no other combination of metric cuts that could consistently remove obvious poor quality phase curves.
This approach was justified by detailed visual inspection of many phase curves and their residuals from which it was clear that phase curves with extreme $G$ and $\sigma_{G}$ were unphysical, and as expected were caused by a combination of poor coverage, bad photometry, and/or unaccounted rotational variation.
\\

The cuts on the primary apparition (which was always in the $o$-filter) for each asteroid were as follows:
\begin{enumerate}[(i)]
    \item Number of data points (after outlier rejection) $\geq50$
    \item $\geq 1$ observation at phase angle $<5\, \si{deg}$
    \item Phase angle range; $\text{max}(\alpha_{\textrm{app}}) - \text{min}(\alpha_{\textrm{app}}$) $\geq 5\, \si{deg}$
    \item Absolute magnitude uncertainty $\sigma_H$(or $\sigma_{H12}$)$<0.05$
    \item Phase parameter uncertainty $\sigma_G(\text{or}\ \sigma_{G12})<0.2$
\end{enumerate}

All apparitions (primary and non primary, $o$ and $c$-filters) had to satisfy the following requirements:
\begin{enumerate}[(i)]
    \item Number of observations (after outlier rejection) $\geq 10$
    \item Phase angle ratio (the ratio of phase angle range of this apparition to the phase angle range of all observations in the same filter for the asteroid) $\geq 0.3$
    \item Standard deviation of residuals $<0.4$
    \item Mean value of residuals $<0.2$
    \item Absolute value of the Pearson correlation coefficient of residuals $<0.25$
    \item Kolmogorov-Smirnov statistic (for normal distribution) $<0.3$
\end{enumerate}

\section{Comparing $HG$ and $H_{12}G_{12}$} \label{appendix:compare_HG_HG12}

\begin{figure}
    \includegraphics{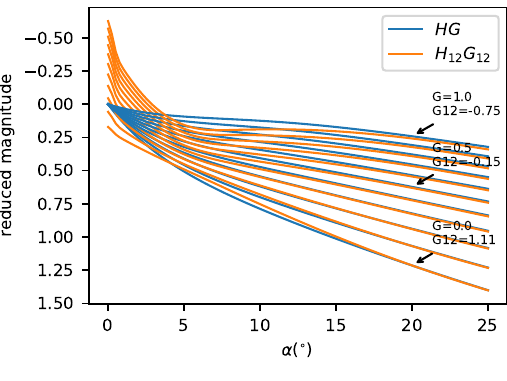}
    \caption{
    Plots of \citetalias{bowellApplicationPhotometricModels1989} model $HG$ phase curves, which were generated with a range of $0\leq G \leq 1$.
    Each $HG$ phase curve was then fitted with the \citetalias{penttilaG1G2Photometric2016} $H_{12}G_{12}$ model.
    The two models have best agreement at large phase angles, however, they vary strongly at lower phase angles leading to differences in absolute magnitude.
    }
    \label{fig:phase_curves_G_G12}
\end{figure}

 In order to compare the properties of the $HG$ \citepalias{bowellApplicationPhotometricModels1989} and $H_{12}G_{12}$ \citepalias{penttilaG1G2Photometric2016} phase curve systems we generated model $HG$ phase curves across a phase angle range of $0\rightarrow25\, \si{deg}$, which is the typical phase angle range for asteroids observed by \ATLAS (figure \ref{fig:phase_angle_max}).
 We generated a number of phase curves with slope parameters $0<G<1$.
 We then performed a linear least squares fit of the $H_{12}G_{12}$ model to each $HG$ phase curve, the results of which are shown in figure \ref{fig:phase_curves_G_G12}.
 Both models are in general agreement at larger phase angles where brightening is approximately linear, however, the differing treatment of the opposition brightness surge (through their basis functions, equations \ref{eqn:B89_model} and \ref{eqn:P16_model}) by the two models is clear as phase angles decreases.
From this simple analysis we obtain an empirical mapping of $G\rightarrow G_{12}$ which provides a good description of the relation between $G$ and $G_{12}$ for the SAP dataset (figure \ref{fig:HG_HG12_relations}).
We performed a 3rd-order polynomial fit for each relation, the parameters of which are as follows 
\begin{subequations}
	\begin{align}
		&G_{12} = -0.7451 G^3 + 2.577 G^2 - 3.716 G + 1.124 \label{eqn:G12_G_polyrelation} \\ 
		&H_{12o} - H_o = -0.1686 G^3 + 0.5697 G^2 - 1.231 G + 0.1815 \label{eqn:H12_G_polyrelation}
	\end{align}
\end{subequations}

This behaviour also explains why our phase curve fits for $H_{12}G_{12}$ more frequently stray into ``unphysical'' solutions, i.e.\ solutions outside of $0\leq G_{12}\leq1$ (figure \ref{fig:phase_curve_G12_P16}).
Furthermore, these model fits clearly illustrate the origin of the different absolute magnitudes obtained with the two phase curve systems (figure \ref{fig:HG_HG12_relations}), where the relation between $\Delta H=H_{12}-H$ and $G$ is reflected well in the \ATLAS data.
This idealised test shows that even if there are observations all the way down to zero phase the fitted $H_{12}G_{12}$ model will differ significantly from the $HG$ fit.
Therefore $\Delta H$ in our dataset is not primarily caused by a lack of information at low phase angles, rather an inherent difference between the two models.

\section{Jupiter Trojan KS bootstrap tests} \label{appendix:Jupiter_Trojan_KS_tests}

\begin{figure*}
\centering
    \includegraphics{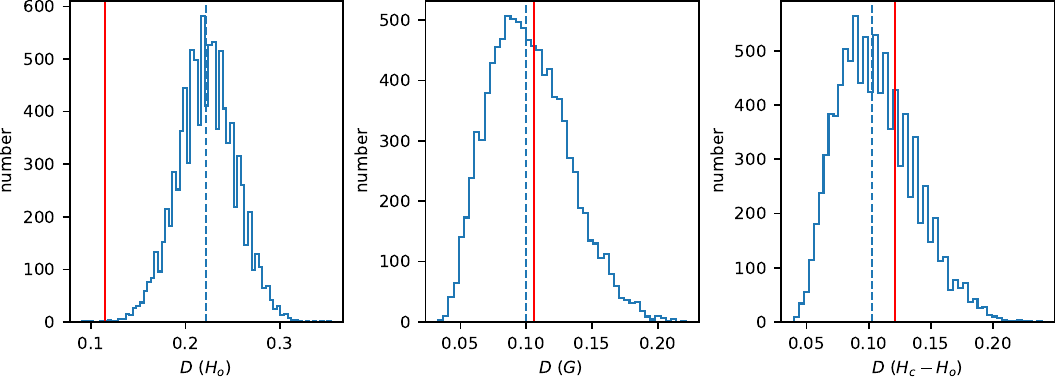}
    \caption{
    Distributions of $D$ values for KS tests comparing $H$, $G$ and $H_c-H_o$ colour of the L4 and L5 Jupiter Trojans in the SAP dataset (left, middle and right panels respectively; compare to figure \ref{fig:jupiter_trojan_L4_L5_KS} CDFs).
    The vertical dashed line indicates the median $D$ of each distribution and the vertical solid line indicates the critical value $D_\text{crit}$ for rejection of the null hypothesis that the two distributions are the same (equation \ref{eqn:KS_D_crit}).
    }
    \label{fig:jupiter_trojan_KS_bootstrap}
\end{figure*}

Here we present the results of performing KS tests on bootstrapped data (resampled with replacement) in order to analyse the distributions of Jupiter Trojan parameters $H$, $G$ and $H_c-H_o$.
For each parameter, 1000 bootstrapped KS tests were performed and the distribution of the resulting $D$ values is shown in figure \ref{fig:jupiter_trojan_KS_bootstrap}.
In these figures we compare the median $D$ value for each set of KS tests and compare this to the critical value of $D$ required for rejecting the null hypothesis at a significance of 0.05.
\begin{equation}
    D_\text{crit} = 1.36 \sqrt{(n+m)/(n\cdot m)}
    \label{eqn:KS_D_crit}
\end{equation}
where $n$ and $m$ are the number of data points in the two samples being considered, and the constant of 1.36 refers to a significance of $5\%$ \citep{knuthArtComputerProgramming1998}. 
As described in section \ref{sec:results-jupiter_trojans} we found that this method was is more accurate than relying on the results of a single KS test of what is a relatively small set of objects.
We confirm differences in the distribution of $H$ between the L4 and L5 groups as observed by \ATLAS; the distribution of $D$ for resampling tests is consistently higher than $D_\text{crit}$ (figure \ref{fig:jupiter_trojan_KS_bootstrap}, left panel).
In contrast, although some tests for slope parameter and colour exceed $D_\text{crit}$, the distributions of $D$ are more often below or near $D_\text{crit}$ (figure \ref{fig:jupiter_trojan_KS_bootstrap}, middle and right panels).
As such this indicates that any differences in $H_c-H_o$ colour and phase parameter $G$ are not statistically significant enough to reject the null hypothesis.

\begin{figure*}
\centering
    \includegraphics{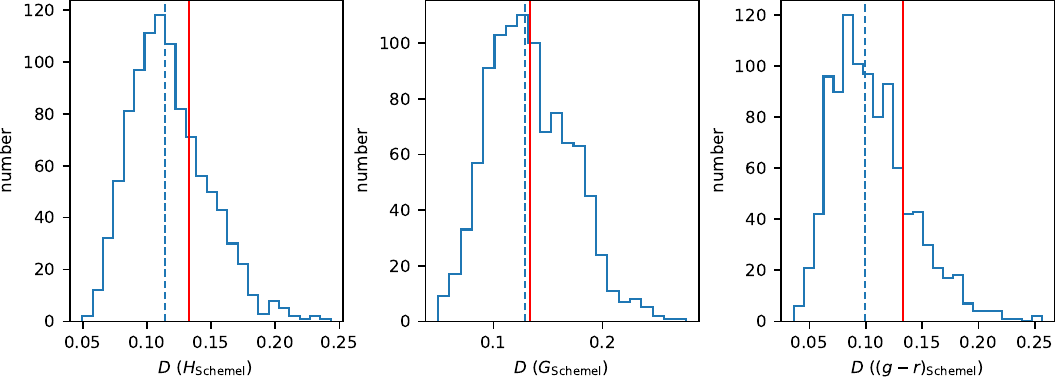}
    \caption{
    Similar to figure \ref{fig:jupiter_trojan_KS_bootstrap}, but analysing the $H$, $G$ and $g-r$ colours of Jupiter Trojans as published by \protect\cite{schemelZwickyTransientFacility2021}.
    }
    \label{fig:schemel_jupiter_trojan_KS_bootstrap}
\end{figure*}

Furthermore, we repeated this analysis for the values of $H$, $G$ and $g-r$ colour published by \citetalias{schemelZwickyTransientFacility2021} (figure \ref{fig:schemel_jupiter_trojan_KS_bootstrap}).
These results show that differences in the distributions of $H$ for the L4/L5 groups is not present in their analysis of ZTF data. 
This is as expected given that the initial \rockAtlas watchlist used in this study started with different $H$ distributions of the L4/L5 Trojans (section \ref{sec:results-jupiter_trojans}).
Similar to our results there are no statistically significant differences between L4/L5 for $G$ or $g-r$ in ZTF data \citepalias{schemelZwickyTransientFacility2021}

\bsp	
\label{lastpage}
\end{document}